\documentclass[sigconf,nonacm]{acmart}

\AtBeginDocument{%
  \providecommand\BibTeX{{%
    \normalfont B\kern-0.5em{\scshape i\kern-0.25em b}\kern-0.8em\TeX}}}

\setcopyright{rightsretained} 
\copyrightyear{2022}
\acmYear{2022}
\acmDOI{10.1145/3545948.3545975}

\acmConference[RAID 2022]{25th International Symposium on Research in Attacks, Intrusions and Defenses}{October 26--28, 2022}{Limassol, Cyprus}
\acmBooktitle{25th International Symposium on Research in Attacks, Intrusions and Defenses (RAID 2022), October 26--28, 2022, Limassol, Cyprus}
\acmISBN{978-1-4503-9704-9/22/10}

\usepackage[normalem]{ulem}
\usepackage{multirow}
\usepackage{pifont}

\usepackage{wasysym}
\usepackage{color}
\usepackage{graphicx}
\usepackage{booktabs}
\usepackage{listings}
\usepackage{parcolumns}
\usepackage{tikz}
\usepackage{float}
\usepackage{subcaption}
\usepackage{circledsteps}
\usepackage{adjustbox}
\usepackage{balance}
\usepackage{enumitem}

\setlist{nosep}

\newcommand{\toolname}{\textsc{Elysium}}
\newcommand{\eg}{e.g.,~}
\newcommand{\ie}{i.e.,~}
\newcommand{\cf}{cf.~}
\newcommand{\etc}{etc.}

\newcommand\mydots{\hbox to 1em{.\hss.\hss.}}
\usetikzlibrary{positioning,arrows,tikzmark,calc}
\definecolor{darkgreen}{HTML}{32CD32}

\usepackage{listings, xcolor}

\definecolor{verylightgray}{rgb}{.97,.97,.97}

\lstdefinelanguage{Solidity}{
	keywords=[1]{anonymous, assembly, assert, balance, break, call, callcode, case, catch, class, constant, continue, contract, debugger, default, delegatecall, delete, do, else, emit, event, export, external, false, finally, for, function, constructor, gas, if, implements, import, in, indexed, instanceof, interface, internal, is, length, library, log0, log1, log2, log3, log4, memory, modifier, new, payable, pragma, private, protected, public, pure, push, require, return, returns, revert, selfdestruct, send, storage, struct, suicide, super, switch, then, this, throw, transfer, true, try, typeof, using, value, view, while, with, addmod, ecrecover, keccak256, mulmod, ripemd160, sha256, sha3}, 
	keywordstyle=[1]\color{blue}\bfseries,
	keywords=[2]{address, bool, byte, bytes, bytes1, bytes2, bytes3, bytes4, bytes5, bytes6, bytes7, bytes8, bytes9, bytes10, bytes11, bytes12, bytes13, bytes14, bytes15, bytes16, bytes17, bytes18, bytes19, bytes20, bytes21, bytes22, bytes23, bytes24, bytes25, bytes26, bytes27, bytes28, bytes29, bytes30, bytes31, bytes32, enum, int, int8, int16, int24, int32, int40, int48, int56, int64, int72, int80, int88, int96, int104, int112, int120, int128, int136, int144, int152, int160, int168, int176, int184, int192, int200, int208, int216, int224, int232, int240, int248, int256, mapping, string, uint, uint8, uint16, uint24, uint32, uint40, uint48, uint56, uint64, uint72, uint80, uint88, uint96, uint104, uint112, uint120, uint128, uint136, uint144, uint152, uint160, uint168, uint176, uint184, uint192, uint200, uint208, uint216, uint224, uint232, uint240, uint248, uint256, var, void, ether, finney, szabo, wei, days, hours, minutes, seconds, weeks, years},	
	keywordstyle=[2]\color{teal}\bfseries,
	keywords=[3]{block, blockhash, acoinbase, difficulty, gaslimit, number, timestamp, msg, data, gas, sender, sig, value, now, tx, gasprice, origin},	
	keywordstyle=[3]\color{violet}\bfseries,
	identifierstyle=\color{black},
	sensitive=false,
	comment=[l]{//},
	morecomment=[s]{/*}{*/},
	commentstyle=\color{gray}\ttfamily,
	stringstyle=\color{red}\ttfamily,
	morestring=[b]',
	morestring=[b]"
}

\begin{document}

\title[\toolname{}: Context-Aware Bytecode-Level Patching to Automatically Heal Vulnerable Smart Contracts]
{\toolname{}: Context-Aware Bytecode-Level Patching to Automatically Heal Vulnerable Smart Contracts}

\author{Christof Ferreira Torres}
\affiliation{%
  \institution{SnT, University of Luxembourg}
  \city{Luxembourg}
  \country{Luxembourg}}
\email{christof.torres@uni.lu}

\author{Hugo Jonker}
\affiliation{%
  \institution{Open University of the Netherlands}
  \city{Heerlen}
  \country{Netherlands}}
\email{hugo.jonker@ou.nl}

\author{Radu State}
\affiliation{%
  \institution{SnT, University of Luxembourg}
  \city{Luxembourg}
  \country{Luxembourg}}
\email{radu.state@uni.lu}

\begin{abstract}
Fixing bugs is easiest by patching source code. However, source code is
not always available: only 0.3\% of the $\sim$49M smart contracts
that are currently deployed on Ethereum have their source code publicly available.
Moreover, since contracts may call functions from other contracts,
security flaws in closed-source contracts may affect open-source contracts as well. 
However, current state-of-the-art approaches that operate on closed-source contracts (\ie EVM
bytecode), such as \textsc{EVMPatch} and \textsc{SmartShield}, make use of purely hard-coded templates that leverage fix patching patterns.
As a result, they cannot dynamically adapt to the bytecode
that is being patched, which severely limits their flexibility and
scalability.
For instance, when patching integer overflows using hard-coded templates, a particular patch template needs to be employed as the bounds to be checked are different for each integer size (\ie one template for \texttt{uint256}, another template for \texttt{uint64}, \etc).

In this paper, we propose \toolname{}, a scalable approach towards automatic smart
contract repair at the bytecode level. \toolname{} combines template-based and
semantic-based patching by inferring context information from bytecode. \toolname{}
is currently able to patch 7 different types of vulnerabilities in smart contracts automatically and can easily
be extended with new templates and new bug-finding tools. We evaluate its effectiveness
and correctness using 3 different datasets by replaying more than 500K transactions on
patched contracts. We find that \toolname{} outperforms existing tools by patching at least 30\% more contracts correctly. 
Finally, we also compare the overhead of \toolname{} in terms of deployment and transaction cost. In comparison to other tools,
we find that generally \toolname{} minimizes the runtime cost (\ie transaction cost) up to a factor of 1.7, for
only a marginally higher deployment cost, where deployment cost is a one-time cost as compared to the runtime cost.
\end{abstract}


\maketitle

\vspace{-0.3cm}
\section{Introduction}

Ideally, bugs in programs should be repaired by simply patching the source code. However,
in some cases, the original source code may not be available. A poignant example
are smart contracts: their bytecode is publicly available via the blockchain,
yet in March 2022, out of 49,183,523 smart contracts deployed on the Ethereum blockchain (via Google BigQuery~\cite{ethereumBigQuery}), only
152,996 (via the Smart Contract Sanctuary project~\cite{smart_contract_sanctuary}) have their source code publicly available. For the remaining $\sim$49M smart contracts,
no source code is publicly available. Moreover, smart contracts may call functions
from other smart contracts. This implies that insecurities in bytecode-only or ``closed-source'' contracts
may even affect contracts for which source code is available. 

There has been prolific research on smart contract security. This includes using ``proxy''
contracts (e.g.,~\cite{zheng2021upgradable,upgradeableopenzeppelin}) to make smart contracts upgradeable, designing clients to automatically detect and block
malicious transactions (e.g.,~\cite{grossman2017online,rodler2018sereum,ferreira2020aegis,chen2020soda,ferreira2021eye}),
as well as building tools to automatically catch bugs prior to deployment using techniques such as symbolic execution
(e.g.,~\cite{Luu2016,nikolic2018finding,mueller2018,torres2018osiris,torres2019art,krupp2018teether}),
abstract interpretation and model checking
(e.g.,~\cite{kalra2018zeus,tsankov2018securify,brent2018vandal,schneidewind2020ethor,frank2020ethbmc}),
or even fuzzing (e.g.,~\cite{jiang2018contractfuzzer,he2019ilf,ferreira2021confuzzius}).
Despite all these efforts, even well-studied bugs with well-known countermeasures
(\eg reentrancy) still occur in high-value contracts.
Prominent examples include the second Parity wallet hack in 2017, where
despite the source code having been manually audited and fixed, an attacker was able to lock up \$150M~\cite{parityhack2}, or the 2020 reentrancy bugs in both the Uniswap and Lendf.me
smart contracts~\cite{uniswaplendfme}, which resulted together in \$25M worth of assets being stolen after being manually audited.
These examples illustrate poignantly that 
automation of both bug finding and bug patching is sorely needed. Most effort has been spent on automating bug discovery and has not carried over to bug patching, which can arguably be seen as one of the main roadblocks to practical smart contract security.

While there has been research into automatically patching smart
contracts~\cite{yu2020smart,zhang2020smartshield,rodler2020evmpatch,nguyen2021sguard},
existing works are still limited: (1) they only address a few types of vulnerabilities, (2) they use hard-coded templates that do not scale (\ie templates are brittle and do not cover all cases, meaning that several templates need to be introduced to cover minor variants of existing cases, such as for example a new template for patching integer overflows with different sizes), and (3) they add a large
overhead in terms of deployment and runtime costs.

We propose a new methodology to address these shortcomings by automatically generating
context-aware patches which adapt to the contract that is being patched. For each contract, we first
perform a number of analyses such as integer type inference and free storage space
inference to understand the context of the smart contract sufficiently to be able
to create tailored and efficient patches. Both analyses and patching are performed
at the bytecode level.
An added bonus is that bytecode level patching results in more efficient code
in terms of size and gas usage than recompiling patched source code~\cite{rodler2020evmpatch}.
We follow a hybrid approach by combining the usability of template-based approaches
with the flexibility and effectiveness of semantic-based approaches. 
Template-based patching simply inserts a fixed sequence of instructions irrespective of the semantics of the program, whereas semantic-based patching modifies the program while preserving its original semantics.
Smart contract developers can either reuse existing patch templates or easily write new patch templates to fix new types of vulnerabilities without having to worry about the context of the smart contract (\eg free state variables, integer types, etc.). Our templates contain place holders which are automatically replaced with contract-related (\ie semantic) information during patch generation.
Moreover, since our approach leverages already existing bug-finding tools, it can
easily be extended to incorporate new bug-finding tools, giving
it the flexibility to handle future vulnerabilities.

\noindent
\textbf{Contributions.} We summarize our contributions as follows:
\begin{itemize}[leftmargin=0.5cm]
    \item We present a novel context-aware bytecode level patching approach that combines template-based with semantic-based patching to create flexible and tailored patches for smart contracts.
    \item We propose \toolname{}, a tool that implements our approach and that is able automatically patche 7 different types of vulnerabilities in smart contracts.
    \item We compare our tool to existing works using 3 different datasets by replaying more than 500K transactions, and demonstrate that \toolname{} not only patches more bugs (at least 30\% more), but also that it is more efficient than existing works in terms of runtime costs (up to 1.7 times less gas). \vspace{0.1cm}
\end{itemize}

\vspace{-0.2cm}
\section{Background}

In this section, we provide background on smart contracts, Ethereum bytecode, and the Ethereum virtual machine.

\vspace{-0.2cm}
\subsection{Smart Contracts}

Ethereum proposes two types of accounts: \textit{externally owned accounts} (EOA) and \textit{contract accounts} (\ie smart contracts). 
Both account types, EOAs and smart contracts, are identifiable via a unique 160-bit address and contain a balance that keeps track of the amount of ether owned by the account.
While EOAs are controlled via private keys and have no associated code, smart contracts are not controlled via private keys and have associated code. 
As a result, smart contracts operate as fully-fledged programs that are stored and executed across the Ethereum blockchain. 
They are different from traditional programs in many ways.
Once deployed, smart contracts cannot be removed or updated, unless they have been explicitly designed to do so. 
Smart contracts are deployed by leaving a transaction's receiver empty and adding the code of the contract to be deployed to a transaction's data field.
After deployment, smart contract functions can be invoked by encoding the function signature and arguments in a transaction’s data field. A so-called \textit{fallback} function is executed whenever the provided function name is not implemented. 
A key-value store allows smart contracts to persist state across transactions.
Smart contracts are usually developed using a high-level programming language.
Despite a plethora of programming languages \cite{vyper,lll,bamboo,coblenz2017obsidian}, Solidity \cite{solidity} remains the most prominent language for developing smart contracts in Ethereum.
Independently of the chosen programming language, the high-level source code must be translated into a low-level representation, so-called Ethereum bytecode, in order to be executable by the Ethereum Virtual Machine (EVM).

\subsection{Ethereum Bytecode}

\begin{figure}
    \centering
    \includegraphics[width=1.0\columnwidth]{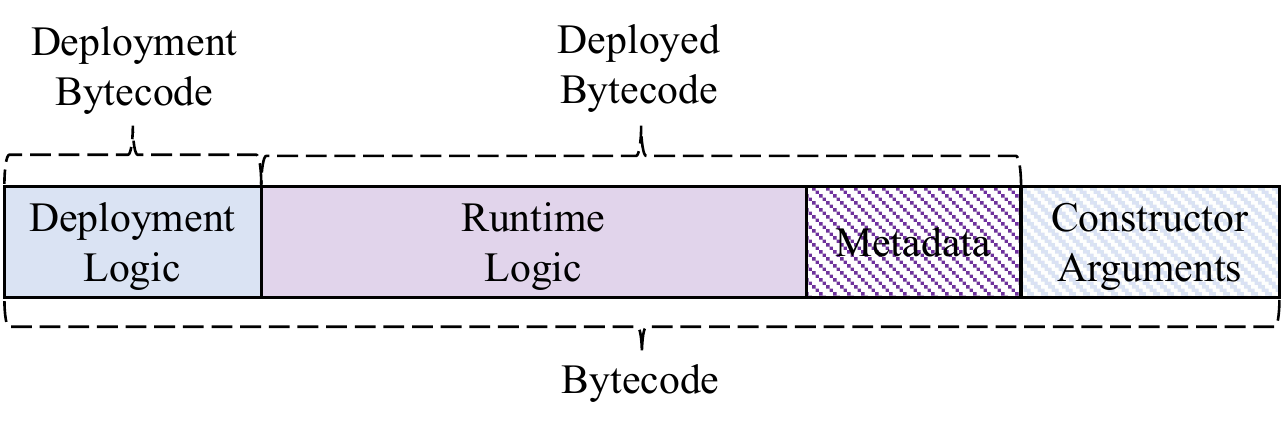}
    \caption{An illustrative example of the anatomy of Ethereum bytecode. Bytecode consists of two main parts: deployment bytecode and deployed bytecode.}
    \label{fig:bytecode_anatomy}
\end{figure}

Ethereum bytecode consists of a sequence of bytes that is interpreted by the EVM. Each byte either encodes an instruction or a byte of data.
\figurename~\ref{fig:bytecode_anatomy} depicts the anatomy of Ethereum bytecode. Ethereum bytecode consists of two main parts: \textit{deployment bytecode} and \textit{deployed bytecode}. 
Deployment bytecode includes the deployment logic of the smart contract. 
This logic is responsible for initializing state variables and reading constructor arguments appended at the end of the Ethereum bytecode. It is also in charge of extracting the deployed bytecode from the Ethereum bytecode and copying it to persistent storage. 
This is achieved via the \texttt{CODECOPY} and \texttt{RETURN} instructions. 
Starting from a given offset and for a given size, the \texttt{CODECOPY} instruction first copies the code running in the current environment to memory.
Afterwards, the \texttt{RETURN} instruction returns the code copied in memory to the EVM. As a result, the EVM creates a new contract by generating a new 160-bit address and persisting the returned code with this address.
The deployed bytecode contains the \textit{runtime logic} (\ie runtime bytecode) and optional \textit{metadata}. 
The runtime bytecode is the logic that is executed whenever a transaction is sent to a smart contract.
Some compilers, such as the Solidity compiler, also append some metadata (\eg compiler version) to the end of the runtime bytecode.

\subsection{Ethereum Virtual Machine}

The EVM is a stack-based, register-less virtual machine
that runs low-level bytecode and supports a Turing-complete set of instructions. 
Every instruction is represented by a one-byte opcode.
The instruction set currently consists of 142 instructions and provides a variety of operations, ranging from basic operations, such as arithmetic operations or control-flow statements, to more specific ones, such as the modification of a contract’s storage or the querying of properties related to the executing transaction (\eg sender) or the current blockchain state (\eg block number). 
The EVM follows the Harvard architecture model by separating code and data into different address spaces. The EVM possesses four different address spaces: an immutable code address space, which contains the smart contract's bytecode, a mutable but persistent storage address space that allows smart contracts to persist their data across executions, a mutable but volatile memory address space that acts as a temporary data storage for smart contracts during execution, and finally a stack address space that allows smart contracts to pass arguments to instructions at runtime.
Moreover, the EVM employs a gas mechanism that assigns a cost to each instruction. This mechanism prevents denial-of-service attacks and ensures termination.
When issuing a transaction, the sender has to specify a gas limit and a gas price. The gas limit is specified in gas units and must be large enough to cover the amount of gas consumed by the instructions during a contract’s execution. Otherwise, the execution will terminate, and its effects will be rolled back. The gas price defines the amount of ether that the sender is willing to pay per unit of gas. 

\subsection{Smart Contract Vulnerabilities}

In the last few years, a plethora of smart contract vulnerabilities have been identified and studied \cite{atzei2017survey,perez2019smart}.
The NCC Group initiated the Decentralized Application Security Project (DASP) with the goal of grouping the most common smart contract vulnerabilities into categories and ranking them based on their real-world impact \cite{dasp}. Table~\ref{tbl:daspranking}, lists the top 5 categories and their associated vulnerabilities.
Although more categories and vulnerabilities exist, our work primarily focuses on the top 5 categories. We leave it to future work to design patch templates for the missing categories.

\begin{table}
  \caption{Decentralized Application Security Project Top 5}
  \label{tbl:daspranking}
  \begin{adjustbox}{width=\columnwidth,center}
  \small
  \begin{tabular}{r p{2.0cm} p{5.0cm}}
    \toprule
    \textbf{Rank} & \textbf{Category} & \textbf{Associated Vulnerabilities} \\
    \midrule
    1 & Reentrancy      & Same- and Cross-Function Reentrancy \\
    2 & Access Control  & Transaction Origin, Suicidal, Leaking, Unsafe Delegatecall \\
    3 & Arithmetic      & Integer Overflows and Underflows \\
    4 & Unchecked Low Level Calls & Unhandled Exceptions \\
    5 & Denial of Services & Unhandled Exceptions, Transaction Origin, Suicidal, Leaking, Unsafe Delegatecall \\
  \bottomrule
\end{tabular}
\end{adjustbox}
\end{table}
\section{Methodology}

In this section, we describe the individual challenges as well as our approach towards patching vulnerabilities at the bytecode level for the vulnerabilities listed in Table~\ref{tbl:daspranking}.

\subsection{Patching Reentrancy Bugs}

\begin{figure}[ht]
\centering
\begin{subfigure}[b]{0.47\textwidth}
\begin{lstlisting}[language=Solidity,escapechar=\%]
mapping (address => uint) public userBalances;
...
function withdrawBalance() public {
    uint amount = userBalances[msg.sender];
    msg.sender.call.value(amount)("");
    userBalances[msg.sender] = 0;
}
\end{lstlisting}
   \caption{Before Patching \vspace{0.3cm}}
   \label{fig:example_reentrancy} 
\end{subfigure}
\begin{subfigure}[b]{0.47\textwidth}
\begin{lstlisting}[language=Solidity,escapechar=\%]
mapping (address => uint) public userBalances;
%\makebox[0pt][l]{\color{darkgreen!20}\rule[-0.35em]{7.95cm}{1.14em}}%+ bool private locked = false;
...
function withdrawBalance() public {
    uint amount = userBalances[msg.sender];
%\makebox[0pt][l]{\color{darkgreen!20}\rule[-0.35em]{7.95cm}{1.14em}}%+   require(!locked);
%\makebox[0pt][l]{\color{darkgreen!20}\rule[-0.35em]{7.95cm}{1.14em}}%+   locked = true;
    msg.sender.call.value(amount)("");
%\makebox[0pt][l]{\color{darkgreen!20}\rule[-0.35em]{7.95cm}{1.14em}}%+   locked = false;
    userBalances[msg.sender] = 0;
}
\end{lstlisting}
   \caption{After Patching}
   \label{fig:example_no_reentrancy}
\end{subfigure}
\caption{(a) Example of a function vulnerable to reentrancy due to an unguarded external call. (b) Example of a function not vulnerable to reentrancy due to a state variable guarding the external call.}
\vspace{-0.3cm}
\end{figure}

\noindent
The code snippet in \figurename~\ref{fig:example_reentrancy} provides an example of a function that is vulnerable to reentrancy at line 5. The function {\small \texttt{withdrawBalance}} transfers the balance of a user to the calling address. Note that a transfer is simply a call to an address. Hence, if {\small \texttt{\textcolor{violet}{\textbf{msg.sender}}}} is a contract, then the transfer will trigger the code that is associated to {\small \texttt{\textcolor{violet}{\textbf{msg.sender}}}}. This code can be malicious and call back the {\small \texttt{withdrawBalance}} function, and reenter function {\small \texttt{withdrawBalance}} while the first invocation has not finished yet. The issue here is that {\small \texttt{userBalances[\textcolor{violet}{\textbf{msg.sender}}]}} has not been set to zero at that moment, and therefore an attacker can repeatedly withdraw its balance from the contract. This is clearly a concurrency issue that can be addressed in several ways. One solution, is to ensure that all state changes, such as the setting of {\small \texttt{userBalances[\textcolor{violet}{\textbf{msg.sender}}]}} to zero, are performed before the call. However, this requires correctly identifying all state variable assignments that are affected by the call, and moving them before the call. Unfortunately, this process is rather tedious and error-prone, as it might break the semantics of a contract. 
A far more simple and less invasive approach, is to make use of \textit{mutual exclusion}, a well studied paradigm from concurrent computing with the purpose of preventing race conditions \cite{dijkstra2001solution}. The idea is to introduce a so-called \textit{mutex} variable that locks the execution state and prevents concurrent access to a given resource. 
\figurename~\ref{fig:example_no_reentrancy} depicts a patched version of the function {\small \texttt{withdrawBalance}} using mutual exclusion. A new state variable called {\small \texttt{locked}} has been introduced at line 2. The variable is used as a mutex variable and is initially set to {\small \texttt{\textcolor{blue}{\textbf{false}}}}.
The condition at line 6 first checks if {\small \texttt{locked}} is set to {\small \texttt{\textcolor{blue}{\textbf{false}}}} before executing the call at line 8.
Then, before executing the call, the variable {\small \texttt{locked}} is set to {\small \texttt{\textcolor{blue}{\textbf{true}}}} and when the call has finished executing, the variable is set back to {\small \texttt{\textcolor{blue}{\textbf{false}}}}. 
This mechanism ensures that the call at line 8 is not re-executed when the function {\small \texttt{withdrawBalance}} is reentered.
Nevertheless, special care needs to be taken when working with mutexes.
One has to make sure that there is no possibility for a lock to be claimed and never released, otherwise a so-called \textit{deadlock} might occur and render the smart contract unusable. 
However, the greatest challenge of this approach is the introduction of a new state variable at the bytecode level.
While this is straightforward when working at the source code level, it becomes more challenging when working at the bytecode level, where high level information such as state variable declarations are missing. 
Our idea is to use bytecode level taint analysis in order to learn about occupied storage space and infer which storage space is still available for inserting a new state variable (\cf Section \ref{sec:desingandimplementation} for more details on free storage space inference).
It is crucial that we only introduce mutex variables at free storage space as otherwise we will overwrite already used storage space and break the semantics of the contract.
Please note that the code presented in \figurename~\ref{fig:example_reentrancy} is an example of a so-called \textit{same-function} reentrancy. However, Rodler et al. \cite{rodler2018sereum} presented other types of reentrancy such as \textit{cross-function} reentrancy, \textit{delegated} reentrancy, and \textit{create-based} reentrancy. The idea is that an attacker can take advantage of a different function that shares the same state with the reentrancy vulnerable function.
Thus, for a contract to be safe against any type of reentrancy, we have to apply the same locking mechanism to every function that shares state with the function that is vulnerable to reentrancy.
We achieve this by searching the bytecode for writes to the same state variable used inside the reentrancy vulnerable function and by guarding them using the same mutex variable that is used in the reentrancy vulnerable function.

\subsection{Patching Access Control Bugs}

Access control bugs includes: transaction origin, suicidal, leaking, and unsafe delegatecall. While the former requires its own approach, the latter three can be patched using a common approach.

\begin{figure}
\centering
\begin{subfigure}[b]{0.47\textwidth}
\begin{lstlisting}[language=Solidity,escapechar=\%]
address public owner;
...
function withdraw(address receiver) public {
    require(tx.origin == owner);
    receiver.transfer(this.balance);
}
\end{lstlisting}
   \caption{Before Patching \vspace{0.3cm}}
   \label{fig:example_transaction_origin} 
\end{subfigure}
\begin{subfigure}[b]{0.47\textwidth}
\begin{lstlisting}[language=Solidity,escapechar=\%]
address public owner;
...
function withdraw(address receiver) public {
%\makebox[0pt][l]{\color{red!20}\rule[-0.35em]{7.95cm}{1.14em}}%-   require(tx.origin == owner);
%\makebox[0pt][l]{\color{darkgreen!20}\rule[-0.35em]{7.95cm}{1.14em}}%+   require(msg.sender == owner);
    receiver.transfer(this.balance);
}
\end{lstlisting}
   \caption{After Patching}
   \label{fig:example_no_transaction_origin}
\end{subfigure}
\caption{(a) Example of a function vulnerable to transaction origin due to the use of {\small \texttt{\textcolor{violet}{\textbf{tx.origin}}}}. (b) Patched example using {\small \texttt{\textcolor{violet}{\textbf{msg.sender}}}} instead of {\small \texttt{\textcolor{violet}{\textbf{tx.origin}}}}.}
\vspace{-0.3cm}
\end{figure}

\vspace{0.2cm}
\noindent
\textbf{\textit{Patching Transaction Origin.}}
The function {\small \texttt{withdraw}} in \figurename~\ref{fig:example_transaction_origin} makes use of {\small \texttt{\textcolor{violet}{\textbf{tx.origin}}}} to check if the calling address is equivalent to the owner.
However, as {\small \texttt{\textcolor{violet}{\textbf{tx.origin}}}} does not return the last calling address but the address that initiated the transaction, an attacker can try to forward a transaction initiated by the owner in order to impersonate itself as the owner and bypass the check at line 4.
The process of patching a transaction origin vulnerability is rather simple.
\figurename~\ref{fig:example_no_transaction_origin} depicts a patched version of the function {\small \texttt{withdraw}}. The patch simply replaces {\small \texttt{\textcolor{violet}{\textbf{tx.origin}}}} with {\small \texttt{\textcolor{violet}{\textbf{msg.sender}}}}, which returns the latest calling address instead of the origin address, therefore not allowing an attacker anymore to impersonate itself as the owner.

\begin{figure}
\centering
\begin{subfigure}[b]{0.47\textwidth}
\begin{lstlisting}[language=Solidity,escapechar=\%]
contract Suicidal {
    ...
    function kill() public {
        selfdestruct(msg.sender);
    }
}
\end{lstlisting}
   \caption{Before Patching \vspace{0.3cm}}
   \label{fig:example_suicidal} 
\end{subfigure}
\begin{subfigure}[b]{0.47\textwidth}
\begin{lstlisting}[language=Solidity,escapechar=\%]
contract NonSuicidal {
%\makebox[0pt][l]{\color{darkgreen!20}\rule[-0.35em]{7.95cm}{1.14em}}%+   address private owner;
    ...
%\makebox[0pt][l]{\color{darkgreen!20}\rule[-0.35em]{7.95cm}{1.14em}}%+   constructor() {
%\makebox[0pt][l]{\color{darkgreen!20}\rule[-0.35em]{7.95cm}{1.14em}}%+       owner = msg.sender;
%\makebox[0pt][l]{\color{darkgreen!20}\rule[-0.35em]{7.95cm}{1.14em}}%+   }
    ...
    function kill() public {
%\makebox[0pt][l]{\color{darkgreen!20}\rule[-0.35em]{7.95cm}{1.14em}}%+       require(msg.sender == owner);
        selfdestruct(msg.sender);
    }
}
\end{lstlisting}
   \caption{After Patching}
   \label{fig:example_non_suicidal}
\end{subfigure}
\caption{(a) Example of a suicidal contract due to an unprotected {\small \texttt{\textcolor{blue}{\textbf{selfdestruct}}}}. (b) Example of a non-suicidal contract due to a protected {\small \texttt{\textcolor{blue}{\textbf{selfdestruct}}}}.}
\vspace{-0.2cm}
\end{figure}

\vspace{0.2cm}
\noindent
\textbf{\textit{Patching Suicidal, Leaking, and Unsafe Delegatecall.}}
The contract in 
\figurename~\ref{fig:example_suicidal} is considered suicidal. The function {\small \texttt{kill}} does not verify the calling address. 
As a result, anyone can destroy the contract. 
The vulnerabilities leaking and unsafe delegatecall are similar, although they relate to contracts that allow anyone to either withdraw ether or control the destination of a delegatecall. These three vulnerabilities share the same issue, namely the unprotected access to a critical operation.
The idea is therefore to add the missing logic that limits the access to a critical operation to only a single entity, for example, the creator of the smart contract.
\figurename~\ref{fig:example_non_suicidal} depicts a patched version of the function {\small \texttt{kill}}.
A new state variable {\small \texttt{owner}} has been added (line 2) as well as a constructor (lines 4-6) in order to initialize the variable {\small \texttt{owner}} during deployment with the address of the contract creator.
Finally, a check has been added at line 9 to verify if {\small \texttt{\textcolor{violet}{\textbf{msg.sender}}}} is equivalent to the address stored in the variable {\small \texttt{owner}}.
Similar to reentrancy, this approach requires the identification of free storage space in order to introduce a new state variable {\small \texttt{owner}}. 
To initialize the variable {\small \texttt{owner}} at deployment (\cf Section \ref{sec:desingandimplementation} for more details on modifying the deployment bytecode), we are required to modify the deployment bytecode instead of the runtime bytecode.
Please note that before creating a new owner variable, we first try to infer and reuse existing owner variables by employing certain heuristics (\eg identify variables where {\small \texttt{\textcolor{violet}{\textbf{msg.sender}}}} is written to).
Also note that, deployment bytecode always contains a constructor at the bytecode level, we therefore just append an assignment to the end of the existing constructor bytecode.

\subsection{Patching Arithmetic Bugs}

\begin{figure}
\centering
\begin{subfigure}[b]{0.47\textwidth}
\begin{lstlisting}[language=Solidity,escapechar=\%]
mapping (address => uint32) public tokens;
...
function buy(uint32 amount) public { 
    require(msg.value == amount);
    tokens[msg.sender] += amount;
}
\end{lstlisting}
   \caption{Before Patching \vspace{0.3cm}}
   \label{fig:example_integer_overflow} 
\end{subfigure}
\begin{subfigure}[b]{0.47\textwidth}
\begin{lstlisting}[language=Solidity,escapechar=\%]
mapping (address => uint32) public tokens;
...
function buy(uint32 amount) public { 
    require(msg.value == amount);
%\makebox[0pt][l]{\color{darkgreen!20}\rule[-0.35em]{7.95cm}{1.14em}}%+   uint32 bounds = 2**32-1 - tokens[msg.sender];
%\makebox[0pt][l]{\color{darkgreen!20}\rule[-0.35em]{7.95cm}{1.14em}}%+   require(bounds >= amount); 
    tokens[msg.sender] += amount;
}
\end{lstlisting}
   \caption{After Patching}
   \label{fig:example_no_integer_overflow}
\end{subfigure}
\caption{(a) Example of a function vulnerable to an integer overflow due to a missing bounds check guarding the update of {\small \texttt{tokens[\textcolor{violet}{\textbf{msg.sender}}]}}. (b) Example of a function not vulnerable to integer overflows due to an added bounds check guarding the update of {\small \texttt{tokens[\textcolor{violet}{\textbf{msg.sender}}]}}.}
\vspace{-0.2cm}
\end{figure}

\noindent
Arithmetic bugs such as integer overflows and underflows are a common issue in smart contracts. In 2018, several ERC-20 token smart contracts have been victims to attacks due to integer overflows \cite{advisories}.
The code snippet in \figurename~\ref{fig:example_integer_overflow}, provides an example of a function that is vulnerable to an integer overflow at line 5. 
The function {\small \texttt{buy}} is missing a check that verifies if the value contained in {\small \texttt{tokens[\textcolor{violet}{\textbf{msg.sender}}]}} would overflow if {\small \texttt{amount}} would be added.
A common way to ensure that unsigned integer operations do not wrap, is to use the \textit{SafeMath} library provided by OpenZeppelin \cite{safemath}. For example, in the case of addition, the library performs a post-condition check, where it first computes the result of $a + b$ and then checks if the result is smaller than $a$. If this is the case, then an overflow has happened and the library halts and reverts the execution.
However, Solidity allows developers to make use of smaller types (\eg {\small \texttt{uint32}}, {\small \texttt{uint16}}, \etc) in order to use less storage space and therefore reduce costs, despite the EVM being able to operate only on 256-bit values.
As a result, the Solidity compiler artificially enforces the wrapping of integers on these smaller types to be consistent with the wrapping performed by the EVM on types of 256-bit.
Unfortunately, the checks provided by the SafeMath library only work with values of type {\small \texttt{uint256}} and do not protect the developers from integer overflows caused by variables of smaller types. Moreover, Solidity enables integer variables to be unsigned or signed, but SafeMath only checks for unsigned integers. 
Existing approaches such as \textsc{EVMPatch} \cite{rodler2020evmpatch}, \textsc{SmartShield} \cite{zhang2020smartshield} and \textsc{sGuard} \cite{nguyen2021sguard}, leverage hard-coded templates, which follow the same limitation as the SafeMatch library, meaning that they are primarily designed to block integer overflows of 256-bit. Developers can write new templates for different integer sizes, but they cannot apply them to existing approaches as they currently do not differentiate between individual integer sizes and always apply the same template.
Therefore, in order to be able to patch any type of integer overflow, we need to be capable of inferring the size and the signedness (\ie signed or unsigned) of an integer variable. While this is trivial when working with source code, it becomes challenging when working with bytecode, where high-level information such as size and signedness are not directly accessible. The idea of our approach is to leverage bytecode level taint analysis in order to infer the size as well as the signedness of integer variables (\cf Section \ref{sec:desingandimplementation} for more details on integer type inference). 
Once the size and the signedness are determined, we can generate a patch that verifies if an arithmetic operation is in bounds with respect to size and signedness.
For example, \figurename~\ref{fig:example_no_integer_overflow} depicts a patched version of the function {\small \texttt{buy}}.
First, we compute the bounds by subtracting the value of {\small \texttt{tokens[\textcolor{violet}{\textbf{msg.sender}}]}} from the largest possible value of an unsigned 32-bit integer (\ie $2^{32}-1$) (line 5). 
Afterwards, we check if {\small \texttt{amount}} is smaller or equal to the computed bounds (line 6). 
If {\small \texttt{amount}} is not within the computed bounds, then we halt and revert the execution. 
Otherwise, the addition at line 7 is considered safe and we continue the execution.

\subsection{Patching Unchecked Low Level Calls Bugs}

\begin{figure}
\centering
\begin{subfigure}[b]{0.47\textwidth}
\begin{lstlisting}[language=Solidity,escapechar=\%]
uint public prize;
address public winner;
bool public claimed = false;
...
function claimPrize() public {
    require(!claimed && msg.sender == winner);
    msg.sender.send(prize);
    claimed = true;
}
\end{lstlisting}
   \caption{Before Patching \vspace{0.3cm}}
   \label{fig:example_unhandled_exception} 
\end{subfigure}
\begin{subfigure}[b]{0.47\textwidth}
\begin{lstlisting}[language=Solidity,escapechar=\%]
uint public prize;
address public winner;
bool public claimed = false;
...
function claimPrize() public {
    require(!claimed && msg.sender == winner);
%\makebox[0pt][l]{\color{red!20}\rule[-0.35em]{7.95cm}{1.14em}}%-   msg.sender.send(prize);
%\makebox[0pt][l]{\color{darkgreen!20}\rule[-0.35em]{7.95cm}{1.14em}}%+   require(msg.sender.send(prize));
    claimed = true;
}
\end{lstlisting}
   \caption{After Patching}
   \label{fig:example_no_unhandled_exception}
\end{subfigure}
\caption{(a) Example of a function vulnerable to an unhandled exception due to a missing return value check on {\small \texttt{\textcolor{blue}{\textbf{send}}}}. (b) Example of a function not vulnerable to unhandled exceptions due to an added return value check on {\small \texttt{\textcolor{blue}{\textbf{send}}}}.
}
\vspace{-0.2cm}
\end{figure}

\begin{figure*}
    \centering
    \includegraphics[width=0.95\textwidth]{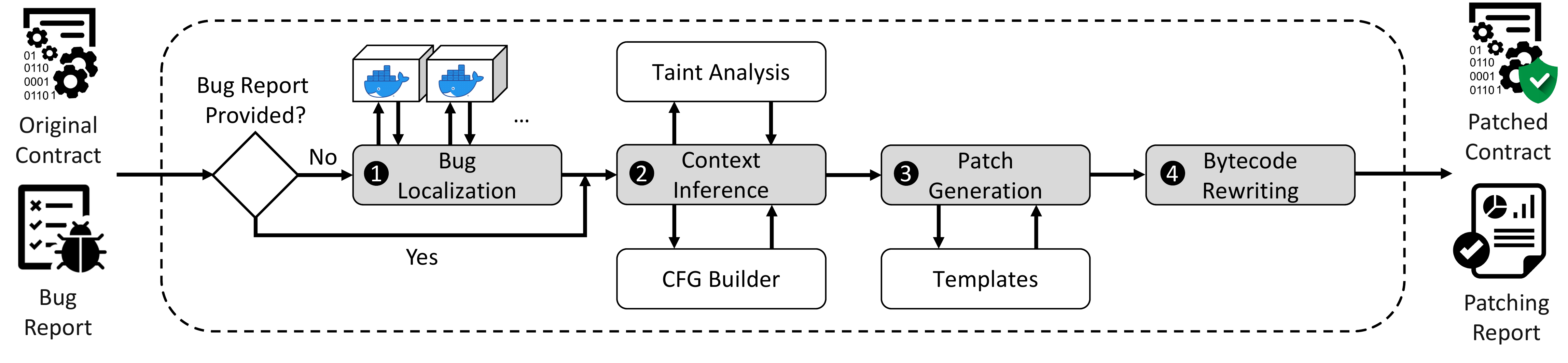}
    \caption{Architecture of \toolname{}. The shaded boxes represent the four main steps of \toolname{}.}
    \label{fig:architecture}
    \vspace{0.5cm}
\end{figure*}

\noindent
An unchecked low level call, also known as an unhandled exception, occurs whenever the return value of a call is not checked. 
A call can fail due to several reasons: an out-of-gas exception, a revert triggered by the called contract, \etc. 
A developer should therefore never assume that a call is always successful, but should always check the return value and handle the case when the call fails.
The function {\small \texttt{claimPrize()}} in \figurename~\ref{fig:example_unhandled_exception} does not check if {\small \texttt{prize}} has been rightfully sent to {\small \texttt{\textcolor{violet}{\textbf{msg.sender}}}} (\cf line 7). As a result, the variable {\small \texttt{claimed}} is set to {\small \texttt{\textcolor{blue}{\textbf{true}}}}, while {\small \texttt{\textcolor{violet}{\textbf{msg.sender}}}} has not received the prize.
Fortunately, patching an unchecked low level call is rather trivial.
A patched version of the function is shown in 
\figurename~\ref{fig:example_no_unhandled_exception}. The patch surrounds the {\small \texttt{\textcolor{blue}{\textbf{send}}}} with a {\small \texttt{\textcolor{blue}{\textbf{require}}}}, which will halt the execution and revert the state in case {\small \texttt{\textcolor{blue}{\textbf{send}}}} is not successful.
Please note that, while this patches the unchecked low level call, the use of {\small \texttt{\textcolor{blue}{\textbf{require}}}} can make in this case the contract vulnerable to denial-of-service attacks if calling {\small \texttt{\textcolor{violet}{\textbf{msg.sender}}}} will always fail.
\section{Design and Implementation}
\label{sec:desingandimplementation}

In this section, we provide details on the overall design and implementation of \toolname{}.

\subsection{Design and Implementation Overview}

\pgfkeys{/csteps/inner color=white}
\pgfkeys{/csteps/fill color=black}

An overview of \toolname{}'s architecture is depicted in \figurename~\ref{fig:architecture}.
\toolname{} takes as input a smart contract as well as an optional bug report and outputs a patched smart contract together with a patching report.
The input smart contract can be either bytecode or Solidity source code. 
The latter, will be compiled into bytecode before performing any analysis or patching.
The patched smart contract consists of the patched version of the bytecode of the original smart contract.
The patching report contains information about execution time and the individual patches that have been applied.
\toolname{}'s patching process follows four main steps: \Circled{\textbf{1}} \textit{bug localization}, \Circled{\textbf{2}} \textit{context inference}, \Circled{\textbf{3}} \textit{patch generation}, and \Circled{\textbf{4}} \textit{bytecode rewriting}.
The bug localization step is responsible for detecting and localizing bugs in the bytecode. This step is skipped in case a bug report is provided.
The context inference step is in charge of building the Control-Flow Graph (CFG) from the byteocde and inferring from the CFG context related information, such as integer types and free storage space, by leveraging taint analysis.
The patch generation step is responsible for creating patches by inserting previously inferred context information within given patching templates.
Finally, as a last step, the bytecode rewriting is in charge of injecting the generated patches into the original CFG and translating it back to bytecode.
\toolname{} is written in Python and consists of roughly 1,600 lines of code\footnote{\toolname{} is publicly available on GitHub: \url{https://github.com/christoftorres/Elysium}.}.
In the following, we describe each of the four steps in more detail.

\subsection{Bug Localization}

In order to be able to patch bugs, \toolname{} first needs to know the exact location of a bug and its type.
One option is to implement our own bug detection solution. However, this is time consuming and error-prone. Another option is to make use of already existing bug detection solutions for smart contracts and to simply incorporate them into \toolname{}. This approach has the advantage of adding modularity by decoupling the detection process from the patching process. This also makes it easy to extend \toolname{} with other or future security analysis tools.
\toolname{} leverages the following three well-known smart contract analysis tools to detect and localize bugs: \textsc{Osiris}\cite{torres2018osiris} to detect integer overflows, \textsc{Oyente}\cite{Luu2016} to detect reentrancy, and \textsc{Mythril}\cite{mueller2018} to detect unhandled exceptions, transaction origin, suicidal contracts, leaking ether, and unsafe delegatecalls.
These tools are provided to \toolname{} as Docker images.
\toolname{} spawns each tool as a separate Docker container, and once a tool has finished running, the output of the tool is parsed and bug information such as the opcode (\eg \texttt{CALL}, \texttt{ADD}), exact bytecode location (\ie program counter) and vulnerability type (\eg reentrancy, integer overflow, \etc) is extracted. This information is then added to a bug report and used by the subsequent steps. Please note that, one can also directly provide a manually crafted bug report to \toolname{}. In such a case, \toolname{} will skip the bug localization step and will directly forward the bug report to the subsequent steps. A user only has to ensure that the bug report follows \toolname{}'s JSON format and that it contains the aforementioned information.

\vspace{-0.2cm}
\subsection{Context Inference}

To effectively patch vulnerabilities related to reentrancy, access control, and integer overflows, we require some context related information.
We gather this information by traversing the CFG and leveraging taint analysis to infer information about integer types and free storage space.
We build the CFG by using the \textit{EVM CFG Builder} python library \cite{evmcfgbuilder}.

\begin{figure*}
    \centering
    \includegraphics[width=0.8\textwidth]{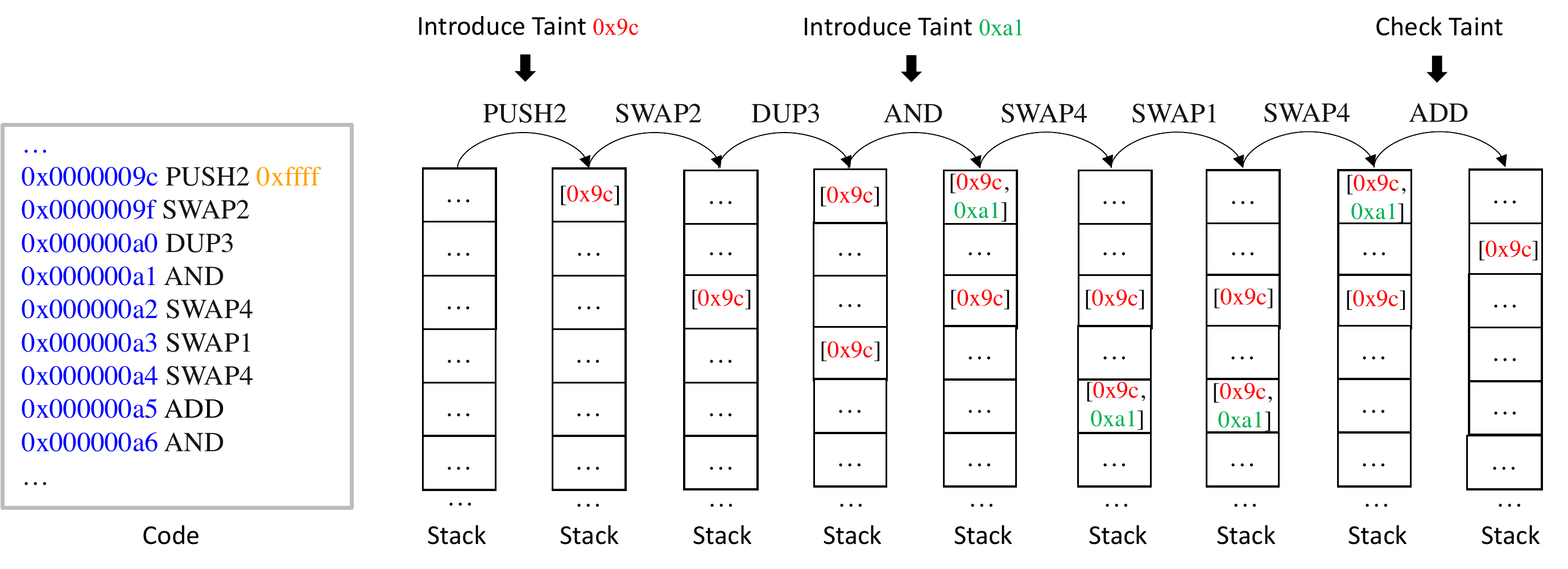}
    \caption{An example on the usage of taint analysis to infer integer types from bytecode.}
    \label{fig:integertypeinference}
    \vspace{0.2cm}
\end{figure*}

\vspace{0.2cm}
\noindent
\textbf{\textit{Integer Type Inference.}}
Integer type information is composed of a size (\eg 32-bit for type {\small \texttt{uint32}})
and a signedness (\eg signed for type {\small \texttt{int}} and unsigned for type {\small \texttt{uint}}).
Both are essential in order to correctly check whether the result of an arithmetic operation is either in-bound or out-of-bound.
However, type information is usually lost during compilation and it is therefore only available at the
source code level. 
Fortunately, we can leverage some behavioral patterns of the Solidity compiler in order to infer the size as well as the signedness of integers.
For example, for unsigned integers, we know that the compiler introduces an \texttt{AND} bitmask in order to “mask off” bits that are not in-bounds with the integer’s size (\ie a zero will mask off the bit, whereas a one will leave the bit set). 
Thus, a variable of type {\small \texttt{uint32}} will result in the compiler adding to the bytecode a \texttt{PUSH} instruction that pushes a bitmask with the value \texttt{0xffffffff} onto the stack followed by an \texttt{AND} instruction.
Hence, from the \texttt{AND} instruction we infer that it is an unsigned integer and from the bitmask
we infer that its size is 32-bit, since \texttt{0xffffffff} = $2^{32} - 1$. 
For signed integers, the compiler
will introduce a sign extension via the \texttt{SIGNEXTEND} instruction. 
A sign extension is the operation of increasing the number of bits of a
binary number while preserving the number’s sign and value. 
The EVM uses two’s complement to represent signed integers.
In two’s complement, a sign extension is achieved by appending ones to the most
significant side of the number. 
The number of ones is computed using $256 - 8(x + 1)$, where $x$ is the first value passed to \texttt{SIGNEXTEND}.
For example, a variable of type {\small \texttt{int32}} will result in the compiler adding to the bytecode a \texttt{PUSH} instruction that pushes the value
$3$ onto the stack followed by a \texttt{SIGNEXTEND}.
Hence, from the \texttt{SIGNEXTEND} instruction we infer that it is a signed integer and from the value $3$ we infer that its size is 32-bit, by solving the following equation: $y = 8(x + 1)$, where in this case $x = 3$. 
Knowing these patterns, we can use taint analysis to infer integer type information at the bytecode level.
First, we iterate in a Breadth First Search (BFS) manner through the CFG until we find the basic block that contains the instruction that is labeled as the bug location.
In the case of integer overflows, the instruction at the bug location can either be an \texttt{ADD}, a \texttt{SUB}, or a \texttt{MUL}.
Afterwards, we use recursion to iterate from the basic block containing the bug back to the root of the CFG, thereby creating along the way a list of all visited instructions. 
This list of instructions reflects the execution path that has to be taken in order to reach the bug location. 
Using this execution path, we can apply taint analysis on it, by executing instruction by instruction and simulating in an abstract manner the effects of each instruction on a shadowed stack, memory, and storage.
The idea is to introduce taint whenever we come across a \texttt{PUSH}, \texttt{AND}, or \texttt{SIGNEXTEND} instruction. Finally, when we arrive at the instruction of the bug location, we check which tainted values have been propagated up to this instruction. For example, if the tainted values that reached the bug location include a \texttt{PUSH} and an \texttt{AND} instruction, then we know that it is an unsigned integer and we know its size from the value introduced by the \texttt{PUSH} instruction.
\figurename~\ref{fig:integertypeinference} provides an illustrative example on how taint is introduced at address \texttt{0x9c} and \texttt{0xa1}, and how it is propagated throughout the stack until it reaches the vulnerable instruction \texttt{ADD} at the address \texttt{0xa6}.

\vspace{0.2cm}
\noindent
\textbf{\textit{Free Storage Space Inference.}}
Patching reentrancy and access control bugs requires the introduction of an additional state variables at the bytecode level.
State variables are associated with EVM storage, a key-value store, where both keys and values are of size 256-bit.
In Solidity, statically-sized variables (\eg everything except mappings and dynamically-sized array types) are laid out contiguously in storage starting from key zero, whereas the storage location for dynamically-sized variables is computed using a hash function.
Moreover, the Solidity compiler tries to pack multiple, contiguous items that need less than 256-bit into a single storage slot.
To not collude with existing statically-sized state variables, we need to find which storage keys are already in use.
To do this, we first extract all the possible execution paths from the CFG by iterating through it in a Depth First Search (DFS) manner and adding each visited instruction to a list. Each list represents one execution path contained in the CFG.
An execution path is terminated whenever we come across a \texttt{STOP}, \texttt{RETURN}, \texttt{SUICIDE}, \texttt{SELFDESTRUCT}, \texttt{REVERT}, \texttt{ASSERTFAIL}, or \texttt{INVALID} instruction. 
Moreover, whenever we run into a \texttt{JUMPI} instruction we split the execution by creating a copy of the list of instructions visited so far and continue iterating first on one branch and then on the other branch.
The EVM provides two different instructions to interact with storage: \texttt{SLOAD} and \texttt{SSTORE}. 
The former takes as input a storage key from the stack and pushes onto the stack the value stored at that key. 
The latter takes as input a storage key and a value, and stores the value at the given key. 
Storage keys are usually pushed onto the stack as constants. 
Thus, whenever a storage instruction is executed (\ie \texttt{SLOAD} or \texttt{SSTORE}), a \texttt{PUSH} instruction will be executed before at some point in the execution with the goal of pushing the storage key onto the stack for the storage instruction to use.
Our idea is therefore to run our taint analysis on all the collected execution paths and to introduce taint whenever we execute a \texttt{PUSH} instruction. 
The taint includes the \texttt{PUSH} instruction and will be propagated across stack as well as memory.
Eventually, we will reach a storage instruction, where we then simply check the taint and infer the used storage key from the propagated \texttt{PUSH} instruction. Afterwards, we add the inferred key to the list of identified storage keys $sk$. Finally, after having analyzed all execution paths, we can compute the next available free storage key as $k = max(sk) + 1$. 
This approach ensures that we do not collude with existing storage keys and it preserves the contiguous layout of state variables in Ethereum smart contract.

\subsection{Patch Generation}

\begin{table*}
  \centering
  \caption{Patch templates provided by \toolname{}.}
  \label{tbl:patchtemplates}
  \begin{tabular}{p{1.9cm} p{8.1cm} p{6.5cm}}
    \toprule
    \textbf{Vulnerability} & \textbf{Patch Template} & \textbf{Source Code Representation} \\
    \midrule
    Reentrancy &
    \small{\texttt{\{\textcolor{violet}{"delete"}: \textcolor{teal}{""}, "\textcolor{violet}{insert}": \textcolor{teal}{"\textcolor{blue}{free\_storage\_location} SLOAD PUSH1\_0x1 EQ ISZERO \textcolor{blue}{PUSH\_jump\_loc\_1} JUMPI PUSH1\_0x1 DUP1 REVERT \textcolor{blue}{JUMPDEST\_jump\_loc\_1} PUSH1\_0x1 \textcolor{blue}{free\_storage\_location} SSTORE"}, \textcolor{violet}{"insert\_mode"}: \textcolor{teal}{"before"}, \textcolor{violet}{"constructor"}: \textcolor{teal}{false}\}}} & \vspace{-1.5em}
    \begin{lstlisting}[language=Solidity,escapechar=\%,frame=none,numbers=none,numbersep=0pt,xleftmargin=0em,framexleftmargin=0em]
%\makebox[0pt][l]{\color{darkgreen!20}\rule[-0.35em]{6.7cm}{1.14em}}%+   require(!locked); 
%\makebox[0pt][l]{\color{darkgreen!20}\rule[-0.35em]{6.7cm}{1.14em}}%+   locked = true; 
\end{lstlisting} \vspace{-2.0em} \\
    & \texttt{...} & \vspace{-0.5em} \texttt{...} \\
    & \small{\texttt{\{\textcolor{violet}{"delete"}: \textcolor{teal}{""}, \textcolor{violet}{"insert"}: \textcolor{teal}{"PUSH1\_0x0 \textcolor{blue}{free\_storage\_location} SSTORE"}, \textcolor{violet}{"insert\_mode"}: \textcolor{teal}{"after"}, \textcolor{violet}{"constructor"}: \textcolor{teal}{false}\}}} & \vspace{-1.5em}
    \begin{lstlisting}[language=Solidity,escapechar=\%,frame=none,numbers=none,numbersep=0pt,xleftmargin=0em,framexleftmargin=0em]
%\makebox[0pt][l]{\color{darkgreen!20}\rule[-0.35em]{6.7cm}{1.14em}}%+   locked = false;
\end{lstlisting} \vspace{-2.0em} \\ \midrule
    Transaction \\ Origin &
    \vspace{-1.8em}
    \small{\texttt{\{\textcolor{violet}{"delete"}: \textcolor{teal}{"ORIGIN"}, \textcolor{violet}{"insert"}: \textcolor{teal}{"CALLER"}, \textcolor{violet}{"insert\_mode"}: \textcolor{teal}{"before"}, \textcolor{violet}{"constructor"}: \textcolor{teal}{false}\}}} & 
    \vspace{-2.8em}
    \begin{lstlisting}[language=Solidity,escapechar=\%,frame=none,numbers=none,numbersep=0pt,xleftmargin=0em,framexleftmargin=0em]
%\makebox[0pt][l]{\color{red!20}\rule[-0.35em]{6.7cm}{1.14em}}%-   require(tx.origin == ...);
%\makebox[0pt][l]{\color{darkgreen!20}\rule[-0.35em]{6.7cm}{1.14em}}%+   require(msg.sender == ...); 
\end{lstlisting} \vspace{-2.0em} \\ \midrule
    Suicidal, \\ Leaking \& \\ Unsafe \\ Delegatecall & \vspace{-4.0em}
    \small{\texttt{\{\textcolor{violet}{"delete"}: \textcolor{teal}{""}, \textcolor{violet}{"insert"}: \textcolor{teal}{"CALLER \textcolor{blue}{free\_storage\_location} SSTORE"}, \textcolor{violet}{"insert\_mode"}: \textcolor{teal}{"after"}, \textcolor{violet}{"constructor"}: \textcolor{teal}{true}\}}} & \vspace{-5.0em}
    \begin{lstlisting}[language=Solidity,escapechar=\%,frame=none,numbers=none,numbersep=0pt,xleftmargin=0em,framexleftmargin=0em]
%\makebox[0pt][l]{\color{darkgreen!20}\rule[-0.35em]{6.7cm}{1.14em}}%+   constructor() {
%\makebox[0pt][l]{\color{darkgreen!20}\rule[-0.35em]{6.7cm}{1.14em}}%+       owner = msg.sender;
%\makebox[0pt][l]{\color{darkgreen!20}\rule[-0.35em]{6.7cm}{1.14em}}%+   } 
\end{lstlisting} \\
    & \vspace{-2.0em} \texttt{...} & \vspace{-2.0em} \texttt{...} \\
    & \vspace{-2.0em} \small{\texttt{\{\textcolor{violet}{"delete"}: \textcolor{teal}{""}, \textcolor{violet}{"insert"}: \textcolor{teal}{"\textcolor{blue}{free\_storage\_location} SLOAD  PUSH20\_0xffffffffffffffffffffffffffffffffffffffff AND CALLER EQ \textcolor{blue}{PUSH\_jump\_loc\_1} JUMPI PUSH1\_0x1 DUP1 REVERT \textcolor{blue}{JUMPDEST\_jump\_loc\_1}"}, \textcolor{violet}{"insert\_mode"}: \textcolor{teal}{"before"}, \textcolor{violet}{"constructor"}: \textcolor{teal}{false}\}}} & \vspace{-2.5em}
    \begin{lstlisting}[language=Solidity,escapechar=\%,frame=none,numbers=none,numbersep=0pt,xleftmargin=0em,framexleftmargin=0em]
%\makebox[0pt][l]{\color{darkgreen!20}\rule[-0.35em]{6.7cm}{1.14em}}%+   require(msg.sender == owner);
\end{lstlisting} \vspace{-2.0em} \\ \midrule
    Integer Overflow (Addition) &
    \small{\texttt{\{\textcolor{violet}{"delete"}: \textcolor{teal}{""}, \textcolor{violet}{"insert"}: \textcolor{teal}{"DUP2 DUP2 \textcolor{blue}{integer\_bounds} SUB LT ISZERO \textcolor{blue}{PUSH\_jump\_loc\_1} JUMPI PUSH1\_0x1 DUP1 REVERT \textcolor{blue}{JUMPDEST\_jump\_loc\_1}"}, \textcolor{violet}{"insert\_mode"}: \textcolor{teal}{"before"}, \textcolor{violet}{"constructor"}: \textcolor{teal}{false}\}}} & 
    \vspace{-1.6em}
    \begin{lstlisting}[language=Solidity,escapechar=\%,frame=none,numbers=none,numbersep=0pt,xleftmargin=0em,framexleftmargin=0em]
%\makebox[0pt][l]{\color{darkgreen!20}\rule[-0.35em]{6.7cm}{1.14em}}%+   require(MAX_VALUE - a >= b); 
\end{lstlisting} \vspace{-2.0em} \\ \midrule
    Integer Overflow (Multiplication) &
    \small{\texttt{\{\textcolor{violet}{"delete"}: \textcolor{teal}{""}, \textcolor{violet}{"insert"}: \textcolor{teal}{"DUP2 DUP2 MUL \textcolor{blue}{integer\_bounds} AND DUP3 ISZERO DUP1 \textcolor{blue}{PUSH\_jump\_loc\_1} JUMPI POP DUP3 SWAP1 DIV DUP2 EQ DUP1 \textcolor{blue}{JUMPDEST\_jump\_loc\_1} SWAP1 POP \textcolor{blue}{PUSH\_jump\_loc\_2} JUMPI PUSH1\_0x1 DUP1 REVERT \textcolor{blue}{JUMPDEST\_jump\_loc\_2}"}, \textcolor{violet}{"insert\_mode"}: \textcolor{teal}{"before"}, \textcolor{violet}{"constructor"}: \textcolor{teal}{false}\}}} & 
    \vspace{-1.6em}
    \begin{lstlisting}[language=Solidity,escapechar=\%,frame=none,numbers=none,numbersep=0pt,xleftmargin=0em,framexleftmargin=0em]
%\makebox[0pt][l]{\color{darkgreen!20}\rule[-0.35em]{6.7cm}{1.14em}}%+   require(b != 0 && a * b / b == a); 
\end{lstlisting} \vspace{-2.0em} \\ \midrule
    Integer Underflow &
    \small{\texttt{\{\textcolor{violet}{"delete"}: \textcolor{teal}{""}, \textcolor{violet}{"insert"}: \textcolor{teal}{"DUP2 DUP2 LT ISZERO \textcolor{blue}{PUSH\_jump\_loc\_1} JUMPI PUSH1\_0x1 DUP1 REVERT \textcolor{blue}{JUMPDEST\_jump\_loc\_1}"}, \textcolor{violet}{"insert\_mode"}: \textcolor{teal}{"before"}, \textcolor{violet}{"constructor"}: \textcolor{teal}{false}\}}} & 
    \vspace{-1.6em}
    \begin{lstlisting}[language=Solidity,escapechar=\%,frame=none,numbers=none,numbersep=0pt,xleftmargin=0em,framexleftmargin=0em]
%\makebox[0pt][l]{\color{darkgreen!20}\rule[-0.35em]{6.7cm}{1.14em}}%+   require(a >= b); 
\end{lstlisting} \vspace{-2.0em} \\ \midrule
    Unhandled \\ Exception &
    \vspace{-1.8em}
    \small{\texttt{\{\textcolor{violet}{"delete"}: \textcolor{teal}{""}, \textcolor{violet}{"insert"}: \textcolor{teal}{"DUP1 ISZERO ISZERO \textcolor{blue}{PUSH\_jump\_loc\_1} JUMPI PUSH1\_0x1 DUP1 REVERT \textcolor{blue}{JUMPDEST\_jump\_loc\_1}"}, \textcolor{violet}{"insert\_mode"}: \textcolor{teal}{"after"}, \textcolor{violet}{"constructor"}: \textcolor{teal}{false}\}}} & 
    \vspace{-2.8em}
    \begin{lstlisting}[language=Solidity,escapechar=\%,frame=none,numbers=none,numbersep=0pt,xleftmargin=0em,framexleftmargin=0em]
%\makebox[0pt][l]{\color{darkgreen!20}\rule[-0.35em]{6.7cm}{1.14em}}%+   require(...); 
\end{lstlisting} \vspace{-2.0em} \\
  \bottomrule
\end{tabular}
\end{table*}

\noindent
To generate patches, we use a combination of template-based and semantic patching. \tablename~\ref{tbl:patchtemplates} provides an overview of all patching templates that are currently provided out-of-the-box by \toolname{}. A patch template is selected according to the vulnerability type that is to be patched. \toolname{} includes templates for seven vulnerability types.
Moreover, existing templates can be modified or new ones can be added by users in order to patch vulnerabilities that are not supported yet by \toolname{}.
We developed our own domain-specific language (DSL) that enables users to easily write and integrate their own context-aware patch templates into \toolname{}.
The structure of a patch template consists of a sequence of instructions to be deleted, a sequence of instructions to be inserted, and an insert mode and constructor flag.
The insert mode  determines whether the instruction sequence to be inserted should be inserted before or after the bug location. The constructor flag specifies if the deletion and insertion should occur at the deployment bytecode.
Our DSL is a combination of the mnemonic representation of EVM instructions and custom keywords that act as place holders for context dependent information.
We leverage the \emph{pyevmasm} library \cite{pyevmasm} to translate the mnemonic representation of EVM instructions into EVM bytecode.
The following four keywords exist:
\texttt{free\_storage\_location},
\texttt{integer\_bounds},
\texttt{PUSH\_jump\_loc\_\{x\}}, and
\texttt{JUMPDEST\_jump\_loc\_\{x\}}.
The \texttt{free\_st
orage\_location} keyword is used to get the current free storage location and it is automatically replaced with a \texttt{PUSH} instruction that pushes the current free storage location onto the stack when generating the patch.
The \texttt{integer\_bounds} keyword is used to get the integer bounds on the instruction at the bug location and it is automatically replaced with a \texttt{PUSH} instruction that pushes the inferred integer bounds onto the stack when generating the patch.
The \texttt{PUSH\_jump\_loc\_\{x\}} and  \texttt{JUMPDEST\_jump\_loc\_\{x\}} keywords work in conjunction. They are used to mark jumps across instructions within a template. The \texttt{PUSH\_jump\_loc\_\{x\}} keyword is replaced in the bytecode rewriting step with a \texttt{PUSH} instruction that pushes the jump address of the \texttt{JUMPDEST\_jump\_loc\_\{x\}} keyword. The  \texttt{JUMPDEST\_jump\_loc\_\{x\}} keyword simply acts as a marker and is afterwards replaced with a normal \texttt{JUMPDEST} instruction.

\vspace{-0.3cm}
\subsection{Bytecode Rewriting}

\begin{figure}
    \centering
    \includegraphics[width=1.0\columnwidth]{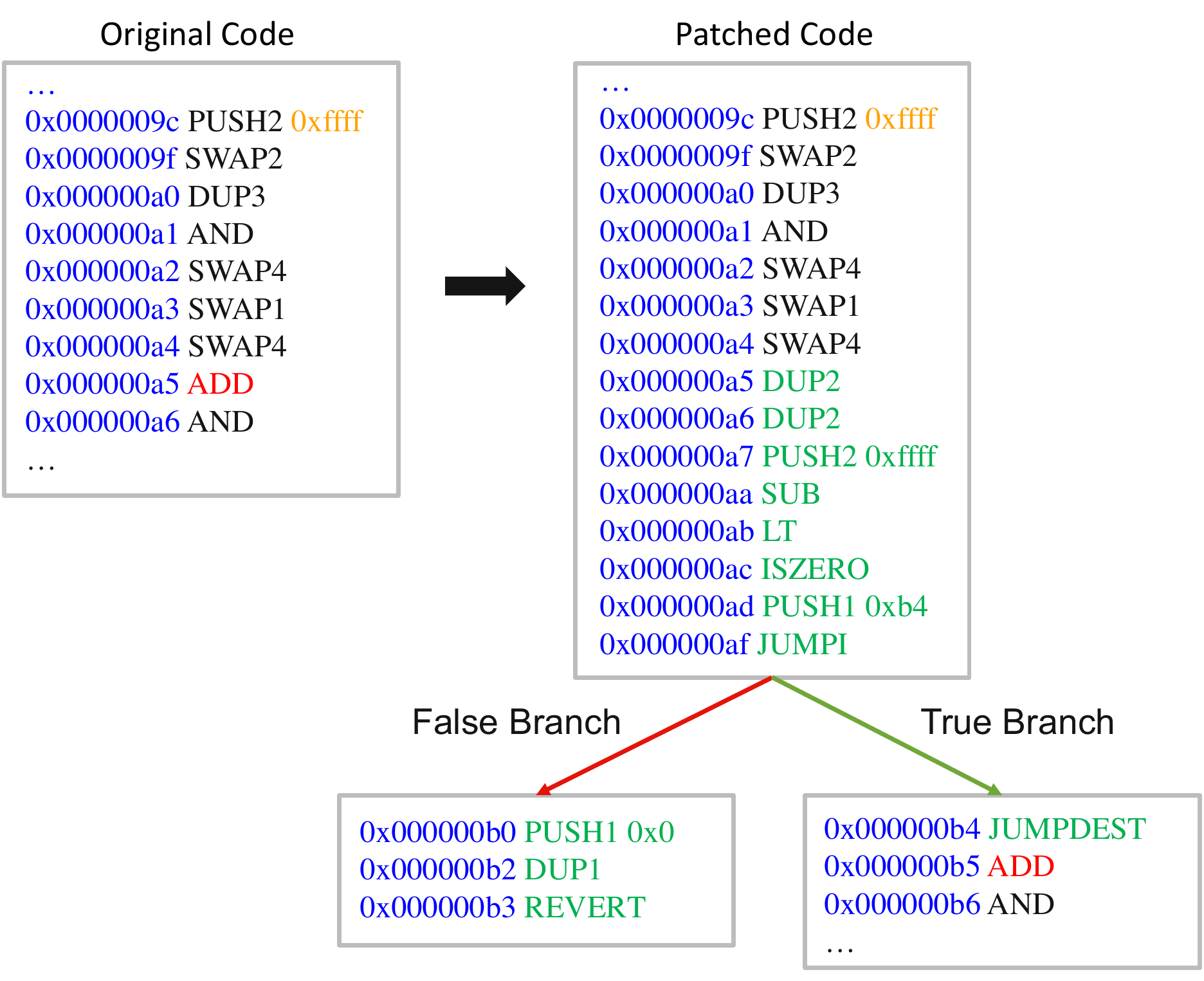}
    \caption{An example on bytecode rewriting, where a guard is added to an unguarded \texttt{ADD} instruction using the \emph{integer overflow (addition)} patch template.}
    \label{fig:bytecoderewriting}
\end{figure}

Ethereum smart contracts are always statically linked, meaning that the bytecode already includes all the necessary library code that is needed at runtime.
This makes EVM bytecode rewriting easier than compared to traditional programs.
Nonetheless, rewriting EVM bytecode still poses some challenges.
Similar to traditional programs, EVM bytecode uses addresses to reference code and data in the bytecode. 
Thus, when modifying the bytecode, one must ensure that the addresses that reference code and data are either adjusted or preserved. 
There are two popular ways to deal with this issue. 
One solution is to preserve the layout of the existing bytecode by copying the basic block that is to be modified at the end of the bytecode. Afterwards, we replace the code of the original basic block with a jump to the copied basic block, and if needed we fill up the original basic block with useless instructions (\eg \texttt{INVALID}, \texttt{JUMPDEST}, \etc) to preserve the original size. 
The modifications are then performed on the copied basic block that resides at the end of the bytecode. At the end of the modified basic block, we jump back to the end of the original basic block such that the rest of the original bytecode can be further executed. This technique is known as "trampoline" and is employed by \textsc{EVMPatch}~\cite{rodler2020evmpatch}. It is the least invasive, since no address references need to be adjusted. 
However, one disadvantage is that the original basic block needs to be large enough to at least hold the logic to jump to the end of the bytecode. Another disadvantage, is the tremendous size increase of the bytecode. While this is less important in traditional programs, for smart contracts this has a monetary impact.
The technique will add useless instructions, so-called "dead code", to preserve the layout, however, this will also result in higher deployment costs. 
As we want to minimize costs, we decided to not employ a trampoline-based approach. Instead,
we opted for a more efficient solution in terms of both deployment and transaction costs, by modifying the bytecode directly at the bug location. However, this technique requires the correct identification of broken address references and the subsequent adjustment according to the new bytecode layout.
Before patching the bytecode, we create a so-called shadow address, a copy of the current address that is associated with each instruction in the CFG.
Then, we scan the CFG for the basic block that is associated with the bug location.
Afterwards,
we modify the basic block by either deleting and/or inserting instructions according to the generated patch.
\figurename~\ref{fig:bytecoderewriting} depicts an example of an original basic block (left hand side) that is vulnerable to an integer overflow at address \texttt{0xa5}, and how it is patched (right hand side) by inserting a patch in the form of a guard ranging from address \texttt{0xa5} to address \texttt{0xb4}. 
After modifying the basic block, we update all the shadow addresses of all instructions in the CFG whose address is larger than the address of the bug location, with the size of the newly added instructions.
For example, for the instruction \texttt{ADD} in \figurename~\ref{fig:bytecoderewriting}, we keep track of the original address with the value \texttt{0xa5} and update the shadow address to the value \texttt{0xb5} (\texttt{0xa5} + $16$ bytes of newly added instructions).
After having patched all the vulnerable basic blocks, we still have to adjust the jump addresses that are pushed onto the stack since some of these might be broken (\eg not reference to a \texttt{JUMPDEST} instruction anymore).
We do this in two steps.
In the first step, we localize broken jump addresses by iterating through each basic block contained in the CFG and scanning each basic block for \texttt{JUMPDEST} instructions where the original address is different than the shadow address. 
In the second step, we iterate through each basic block contained in the CFG and scan each basic block for \texttt{PUSH} instructions whose push value is equivalent to the original address and replace the push value with the shadow address.
Finally, we convert the patched CFG back to bytecode, by first sorting the basic blocks in ascending order according to their starting, and then translating each EVM instruction within the basic block to their bytecode representation.
However, remember that the deployment bytecode copies during deployment the entire runtime bytecode of the smart contract into memory. 
Thus, as the size of the runtime bytecode has changed, the deployment bytecode also needs to be adapted to copy the new amount of runtime bytecode.
We do so by scanning the deployment bytecode for the following consecutive sequence of instructions:
\texttt{PUSH DUP1 PUSH PUSH CODECOPY}. The first \texttt{PUSH} instruction determines the amount of bytes to be copied, the second \texttt{PUSH} instruction determines the offset from where the bytes should be copied, and the third \texttt{PUSH} instruction determines to which offset destination in memory the bytes should be copied.
We update the deployment bytecode by replacing the value of the first \texttt{PUSH} instruction with the new size of the runtime bytecode. The second \texttt{PUSH} instruction is only updated if the deployment bytecode has also been patched (\eg constructor code has been added as part of a patch template).

\begin{table*}
\begin{center}
  \setlength{\tabcolsep}{5.7pt}
  \caption{A comparison of the individual patching tools evaluated in this work.}
  \label{tbl:tools}
  \small
  \begin{tabular}{l l l l l c c c c c c c}
    \toprule
     & & & & & \multicolumn{7}{c}{\textbf{Vulnerabilities}} \\ \cmidrule(lr){6-12}
    \textbf{Toolname} & \textbf{Bug Localization} & \textbf{Patching Level} & \textbf{Approach} & \textbf{Availability} & \textbf{IO} & \textbf{RE} & \textbf{UE} & \textbf{TO} & \textbf{SU} & \textbf{LE} & \textbf{UD} \\
    \midrule
    \textsc{EVMPatch}~\cite{rodler2020evmpatch} & Outsourced & Bytecode & Template & Not Available    &
    $\RIGHTcircle$ & $\Circle$ & $\Circle$ & $\Circle$ & $\LEFTcircle$ & $\LEFTcircle$ & $\LEFTcircle$ \\
    \textsc{SMARTSHIELD}~\cite{zhang2020smartshield} & Outsourced & Bytecode & Template/Semantics & On Request    &
    $\RIGHTcircle$ & $\RIGHTcircle$ & $\CIRCLE$ & $\Circle$ & $\Circle$ & $\Circle$ & $\Circle$ \\
    \textsc{SCRepair}~\cite{yu2020smart} & Outsourced & Source Code & Mutation & Open Source$^\dag$   &
    $\RIGHTcircle$ & $\RIGHTcircle$ & $\CIRCLE$ & $\Circle$ & $\Circle$ & $\Circle$ & $\Circle$ \\
    \textsc{sGuard}~\cite{nguyen2021sguard} & Insourced & Source Code & Template/Semantics & Open Source   &
    $\RIGHTcircle$ & $\CIRCLE$ & $\Circle$ & $\CIRCLE$ & $\Circle$ & $\Circle$ & $\Circle$ \\
    \textbf{\textsc{\toolname}} & \textbf{Outsourced} & \textbf{Bytecode} & \textbf{Template/Semantics} & \textbf{Open Source}   &
    $\CIRCLE$ & $\CIRCLE$ & $\CIRCLE$ & $\CIRCLE$ & $\CIRCLE$ & $\CIRCLE$ & $\CIRCLE$ \\
  \bottomrule
\end{tabular}
\end{center}
\small
$^\dag$ Publicly available source code does not compile. $\Circle$ Not supported. $\RIGHTcircle$ Patching partially supported. $\LEFTcircle$ Patch template must be specified manually. 
$\CIRCLE$ Fully automatic patching supported. \textbf{IO}: integer overflow, \textbf{RE}: reentrancy, \textbf{UE}: unhandled exception, \textbf{SU}: suicidal, \textbf{LE}: leaking, \textbf{UD}: unsafe delegatecall.
\end{table*}

\begin{table}[t]
  \centering
  \caption{CVE dataset overview.}
  \label{tbl:cvedataset}
  \small
  \begin{tabular}{c c r r r r}
    \toprule
    & & & \multicolumn{3}{c}{\textbf{Transactions}} \\ 
    \cmidrule(lr){4-6}
    \textbf{Contract} & \textbf{CVE} & \textbf{Bugs} & \textbf{Total} & \textbf{Benign} & \textbf{Attacks} \\
    \midrule
    BEC & 2018-10299 & 1 & 409,837 & 409,836 & 1 \\
    SMT & 2018-10376 & 1 & 34,164  & 34,163 &  1 \\
    UET & 2018-10468 & 8 & 23,725  & 23,670 & 55 \\
    SCA & 2018-10706 & 9 & 281     &    280 &  1 \\
    HXG & 2018-11239 & 4 & 1,284   &  1,274 & 10 \\
  \bottomrule
\end{tabular}
\end{table}

\begin{table}[t]
  \centering
  \setlength{\tabcolsep}{5pt}
  \caption{\textsc{SmartBugs} dataset overview.}
  \label{tbl:smartbugsdataset}
  \begin{adjustbox}{width=\columnwidth,center}
  \small
  \begin{tabular}{l r r r r}
    \toprule
    & & \multicolumn{3}{c}{\textbf{Bugs}} \\ \cmidrule(lr){3-5}
    
    \textbf{Vulnerability} & \textbf{Contracts} & \textbf{Annotated} & \textbf{Detected} & \textbf{Overlap} \\
    \midrule
    Reentrancy          &  31 &  32 & 29 & \textbf{28} \\
    Access Control      &  18 &  19 & 12 & \textbf{12} \\
    Integer Overflow    &  15 &  23 & 20 & \textbf{16} \\
    Unhandled Exception &  52 &  75 & 23 & \textbf{23} \\
    \midrule
    \textbf{Total}      & \textbf{116} & \textbf{149} & \textbf{84} & \textbf{79} \\ 
  \bottomrule
\end{tabular}
\end{adjustbox}
\end{table}

\begin{table}[t!]
  \centering
  \caption{\textsc{Horus} dataset overview.}
  \label{tbl:horusdataset}
  \begin{adjustbox}{width=\columnwidth,center}
  \begin{tabular}{l r r r r r}
    \toprule
    & & & \multicolumn{3}{c}{\textbf{Transactions}} \\ \cmidrule(lr){4-6}
    \textbf{Vulnerability} & \textbf{Contracts} &  \textbf{Bugs} & \textbf{Total} & \textbf{Benign} & \textbf{Attacks} \\
    \midrule
    Reentrancy              &    44 &    47 & 4,593 &  2,656 & 1,937 \\
    Access Control          &   589 &   823 &  2,116 &    264 & 1,852 \\
    -- Parity Wallet Hack 1 &   585 &   585 &  1,877 &    263 & 1,614 \\
    -- Parity Wallet Hack 2 &   238 &   238 &    358 &    120 &   238 \\
    Integer Overflow        &   125 &   235 & 42,768 & 42,327 &   441 \\
    Unhandled Exception     &   901 &   993 & 80,997 & 78,144 & 2,853 \\
    \midrule
    \textbf{Total Unique}   & \textbf{1,655} & \textbf{2,098} & \textbf{129,863} & \textbf{122,830} & \textbf{7,041} \\ 
  \bottomrule
\end{tabular}
\end{adjustbox}
\end{table}

\section{Evaluation}

In this section, we evaluate \toolname{} by
measuring its effectiveness (RQ1), correctness (RQ2), and costs (RQ3).

\subsection{Experimental Setup}

\vspace{0.2cm}
\noindent
\textbf{\textit{Baselines.}}
We compare \toolname{} to the tools listed in Table~\ref{tbl:tools}. 
Most tools, including \toolname{}, have their bug localization outsourced, meaning that they leverage existing security analysis tools to detect and localize bugs. \textsc{sGuard} is the only tool that leverages its own bug localization.
While \toolname{},
\textsc{EVMPatch}, and \textsc{SmartShield} insert their patches at the bytecode level, other tools such as \textsc{SCRepair} and \textsc{sGuard} insert their patches at the source code level.
Almost all tools, except for \textsc{SCRepair}, use a template-based approach to introduce their patches. However, some tools such as \toolname{}, \textsc{SmartShield}, and \textsc{sGuard} use a combination of template-based and semantic-aware patching. 
The source code of \textsc{EVMPatch} is not publicly available. Nonetheless, the authors released a public dataset with their results for comparison \cite{evmpatchdataset}. \textsc{SmartShield} is only available upon request.
While the source code of \textsc{SCRepair} is publicly available, we did not manage to compile it. Both, \toolname{} and \textsc{sGuard}, are (will be) publicly available under an open source license. 
None of the aforementioned tools, except \toolname{}, are able to patch all the vulnerabilities mentioned in this paper. 
For example, while \textsc{SmartShield} and \textsc{SCRepair} provide means to patch integer overflows, reentrancy, and unhandled exceptions, they do not provide means to patch access control related bugs such as transaction origin or unsafe delegatecall.
Moreover, some tools only provide partial patching capabilities for a given type of vulnerability. For instance, all tools, except \toolname{}, only support the patching of 256-bit unsigned integers and do not support integers of smaller size. Another example is reentrancy, where tools such as \textsc{SmartShield} and \textsc{SCRepair} only provide support for patching same-function reentrancy.
Furthermore, some tools such as \textsc{EVMPatch} require developers to write contract specific patches for access control related bugs and therefore 
do not provide generic fully automatic patching.
\toolname{} on the other hand, provides complete support and fully automatic patching for all vulnerabilities.

\vspace{0.2cm}
\noindent
\textbf{\textit{Datasets.}}
We run our experiments on three different datasets. The first dataset is the CVE dataset \cite{evmpatchdataset} used by Rodler et al. We chose this dataset to be able to compare our tool with \textsc{EVMPatch}. It consists of real-world ERC-20 token contracts that were victims of integer overflow attacks. Moreover, the dataset also provides a list of attacking and benign transactions (see \tablename~\ref{tbl:cvedataset}). However, the dataset is limited to integer overflows and only contains 5 contracts.
The second dataset is the \textsc{SmartBugs} dataset \cite{smartbugs}.
This dataset consists of 116 manually crafted contracts with 149 annotated vulnerabilities across 4 different vulnerability types (see \tablename~\ref{tbl:smartbugsdataset}). While the dataset brings in a large diversity of vulnerabilities, it does not contain a list of benign or attacking transactions. 
The third dataset is the \textsc{Horus} dataset \cite{horus}. The dataset consists of 1,655 unique real-world contracts vulnerable to one of 4 different vulnerabilities, with 129,863 annotated transactions, where 122,830 transactions are benign and 7,041 transactions are attacks (see \tablename~\ref{tbl:horusdataset}).

\subsection{Experimental Results}

\vspace{0.2cm}
\noindent
\textbf{\textit{RQ1: Effectiveness.}}
We first measure the effectiveness of \toolname{} and the other tools on the \textsc{SmartBugs} dataset. The dataset only consists of annotated contracts and does not contain attacking nor benign transactions.
We therefore first run the bug-finding tools (\ie \textsc{Osiris}, \textsc{Oyente}, and \textsc{Mythril}) on the contracts and match the reported bugs with the annotated bugs. The overlap marks the validated ground truth (see overlap column in \tablename~\ref{tbl:smartbugsdataset}).
From the 149 annotated bugs, only 79 overlap with the bugs detected by the bug-finding tools. Moreover, 5 false positives have been reported by the bug-finding tools.
Next, we patch the contracts by running the patching tools and rerunning the bug-finding tools on the patched version returned by each patching tool, and mark a bug as successfully patched if the bug-finding tool does not report the bug anymore.
\tablename~\ref{tbl:smartbugseffectiveness} shows that \toolname{} is able to patch all 79 bugs, whereas \textsc{SmartShield} and \textsc{sGuard} can only patch 45 and 33, respectively. 
We see that \textsc{SmartShield} has issues in patching reentrancy. This is because \textsc{SmartShield} patches reentrancy by moving storage instruction before the call instruction. However, this process is often very imprecise. \textsc{sGuard} has issues in patching integer overflows due to its in-house bug detection not always being able to identify arithmetic bugs correctly.
If we only consider the bug types that all three tools have in common (\ie reentrancy and integer overflows), then we count a total of 23, 31, and 44 bugs patched, for \textsc{SmartShield}, \textsc{sGuard}, and \toolname{}, respectively. This means that \toolname{} patches at least 30\% more bugs than the other tools. 
To measure the effectiveness of \toolname{} and the other tools on the CVE and \textsc{Horus} datasets, we re-execute the attack transactions of each dataset, once on the original bytecode and once one the patched bytecode returned by each tool. We mark an attack as successfully blocked if the patched bytecode resulted in the transaction being reverted.
\textsc{EVMPatch}, \textsc{SmartShield}, and \textsc{Elysium} were able to successfully blocked all attacks for all the contracts within the CVE dataset.
For the \textsc{Horus} dataset, \tablename~\ref{tbl:horusbenignandattacks} shows that \toolname{} is able to successfully block 100\% of all attacks, while \textsc{SmartShield} is able to block only 44\% of all attacks.

\begin{table}[t]
  \centering
  \setlength{\tabcolsep}{4pt}
  \caption{Number of bugs patched by each tool on contracts from the \textsc{SmartBugs} dataset.}
  \label{tbl:smartbugseffectiveness}
  \small
  \begin{tabular}{l r r r r}
    \toprule
    \textbf{Vulnerability} & \textbf{Bugs} & \multicolumn{1}{c}{\textbf{\textsc{SmartShield}}} & \multicolumn{1}{c}{\textbf{\textsc{sGuard}}} & \multicolumn{1}{c}{\textbf{\toolname}} \\
    \midrule
    Reentrancy          & 28 &  7 & 28 & \textbf{28} \\
    Access Control      & 12 & -- &  2 & \textbf{12} \\
    Integer Overflow    & 16 & 16 &  3 & \textbf{16} \\
    Unhandled Exception & 23 & 22 &  -- & \textbf{23} \\
    \midrule
    \textbf{Total} & \textbf{79} & \textbf{45} & \textbf{33} & \textbf{79} \\
  \bottomrule
\end{tabular}
\end{table}

\begin{table}[t]
  \centering
  \setlength{\tabcolsep}{0.5pt}
  \caption{Number of non-blocked benign transactions and blocked attacking transactions from the \textsc{Horus} dataset.}
  \label{tbl:horusbenignandattacks}
  \begin{adjustbox}{width=\columnwidth,center}
  \small
  \begin{tabular}{l r r r r}
    \toprule
    & \multicolumn{2}{c}{\textbf{Non-Blocked Benign}} & \multicolumn{2}{c}{\textbf{Blocked Attacks}} \\
    \cmidrule{2-3} \cmidrule(l){4-5} 
    \textbf{Vulnerability} & \multicolumn{1}{c}{\textbf{\textsc{SmartShield}}} & \multicolumn{1}{c}{\textbf{\toolname}} & \multicolumn{1}{c}{\textbf{\textsc{SmartShield}}} & \multicolumn{1}{c}{\textbf{\toolname}} \\
    \midrule
    Reentrancy                          &  653 (34\%) \hspace{0.1cm} & \textbf{2,608 (98\%)}  & 33 (2\%) \hspace{0.1cm} & \textbf{1,937 (100\%)} \\
    Access Control                      & - \hspace{0.8cm} & \textbf{264 (100\%)} & - \hspace{0.8cm} & \textbf{1,852 (100\%)} \\
    -- Parity Wallet Hack 1             & - \hspace{0.8cm} & \textbf{263 (100\%)} & - \hspace{0.8cm} & \textbf{1,614 (100\%)} \\
    -- Parity Wallet Hack 2             & - \hspace{0.8cm} & \textbf{120 (100\%)} & - \hspace{0.8cm} & \textbf{238 (100\%)} \\
    Integer Overflow                    & 40,996 (97\%) \hspace{0.1cm} & \textbf{41,012 (97\%)} &  432 (98\%) \hspace{0.1cm} & \textbf{441 (100\%)} \\
    Unhandled Exception                 & 63,199 (81\%) \hspace{0.1cm} & \textbf{74,379 (95\%)} & 2,650 (93\%) \hspace{0.1cm} & \textbf{2,853 (100\%)} \\
    \midrule 
    \textbf{Total Unique} & \textbf{104,292 (85\%)} \hspace{0.1cm} & \textbf{117,713 (96\%)} & \textbf{3,073 (44\%)} \hspace{0.1cm} & \textbf{7,041 (100\%)} \\
  \bottomrule
\end{tabular}
\end{adjustbox}
\end{table}

\vspace{0.2cm}
\noindent
\textbf{\textit{RQ2: Correctness.}}
\toolname{}'s correctness depends heavily on the accurate recovery of the CFG and the accurate inference of free storage locations. We downloaded from Etherscan the bytecode and source code for the top 100 smart contracts according to their ether balance. Their lines of source code range from 19 to 3,299 and their number of functions range from 1 to 291. The \textit{EVM CFG Builder} library \cite{evmcfgbuilder} is able to recover 100\% of the CFG for 85 contracts. Overall, the library achieves an average of 96\% recovery with an average time of 6.7 seconds. We improved the library by adding the techniques proposed in \cite{contro2021ethersolve}. The improved version is able to fully recover the CFG for 88 contracts and achieves an average of 98\% recovery with an average time of 7.5 seconds.
For the 12 non-fully recovered contracts, our improved version of the EVM CFG Builder library is able to recover on average 16\% more of the CFG than the original version. To measure the accuracy of the free storage location inference employed by \toolname{}, we leveraged the ability of the Solidity compiler to generate the storage layout of a smart contract and compared the storage layout generated by the Solidity compiler with the storage layout inferred by \toolname{}. \toolname{} is able to correctly infer the storage layout and thus next available free storage location for all 100 contracts.
Besides measuring CFG recovery and free storage location inference, we also measured the correctness of \toolname{} by replaying benign transactions on the patched contracts of the \textsc{EVMPatch} and \textsc{Horus} datasets. A benign transaction is considering successful if the result is identical to the result of the original unpatched transaction.
On the \textsc{EVMPatch} dataset, \textsc{EVMPatch}, \textsc{SmartShield}, and \textsc{Elysium} correctly executed the same number of benign transactions. 
For the \textsc{Horus} dataset, \toolname{} is able to correctly execute 96\% of the benign transactions whereas \textsc{SmartShield} is able to execute only 85\% of the benign transactions (see \tablename~\ref{tbl:horusbenignandattacks}).
We found that the reason for certain benign transactions not being executed correctly is either due to not enough gas being provided for the transaction to be executed on the newly patched smart contract or the CFG simply not being recovered to 100\% and therefore introducing invalid jump destinations that lead to exceptions at runtime.

\captionsetup[figure]{skip=0pt}
\begin{figure}[t]
    \centering
    \includegraphics[width=1.0\columnwidth]{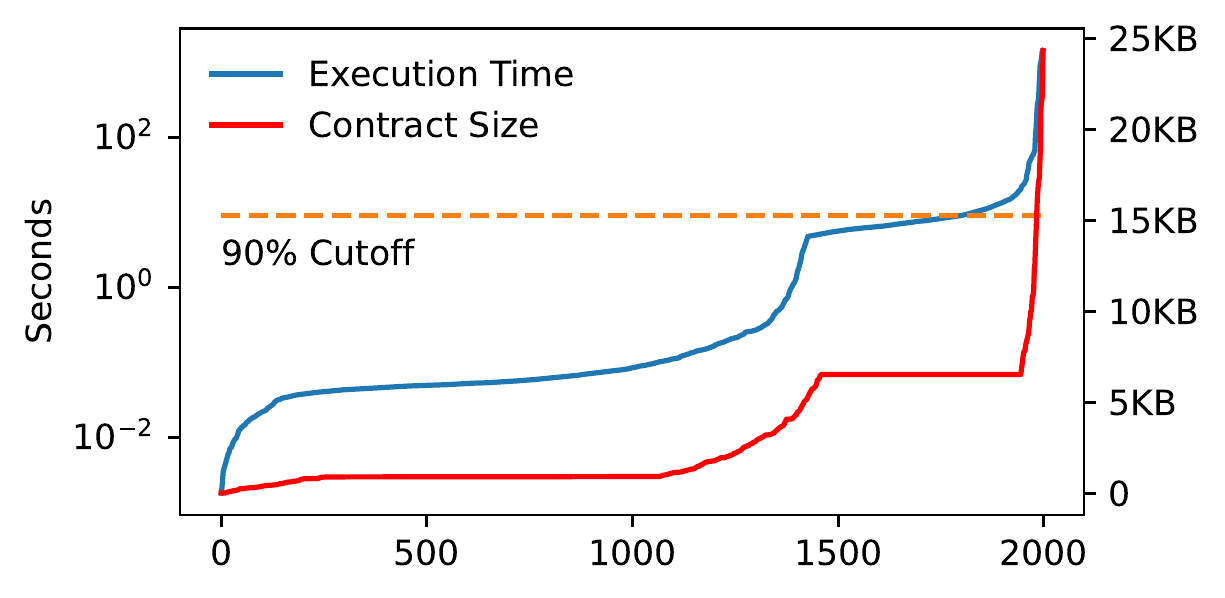}
    \caption{Execution time of \toolname{} on the \textsc{Horus} dataset.}
    \label{fig:executiontime}
\end{figure}
\captionsetup[figure]{skip=10pt}

\begin{figure*}
     \centering
     \begin{subfigure}[b]{0.39\textwidth}
         \centering
         \includegraphics[width=\textwidth]{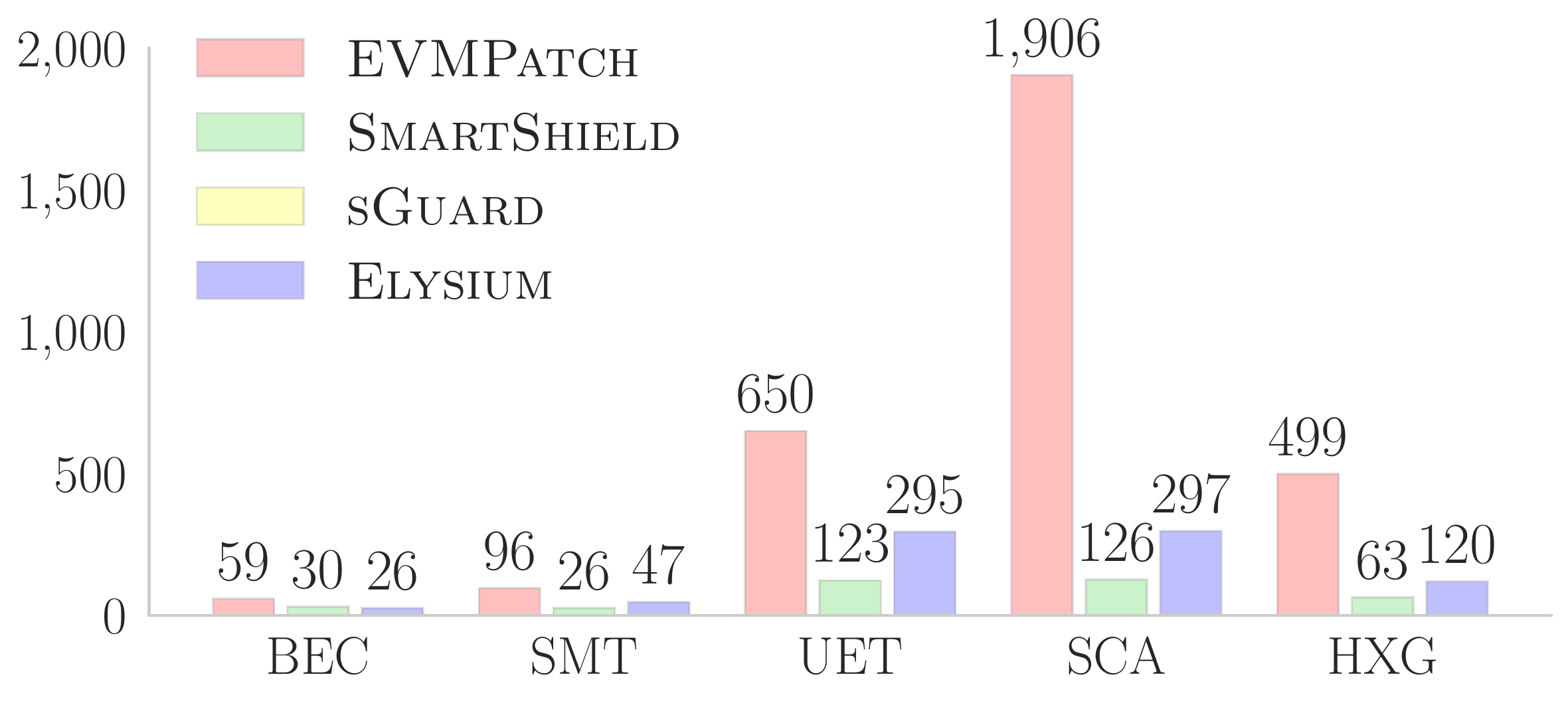}
         \caption{CVE}
         \label{fig:cvedeploymentcosts}
     \end{subfigure}
     \hfill
     \begin{subfigure}[b]{0.29\textwidth}
         \centering
         \includegraphics[width=\textwidth]{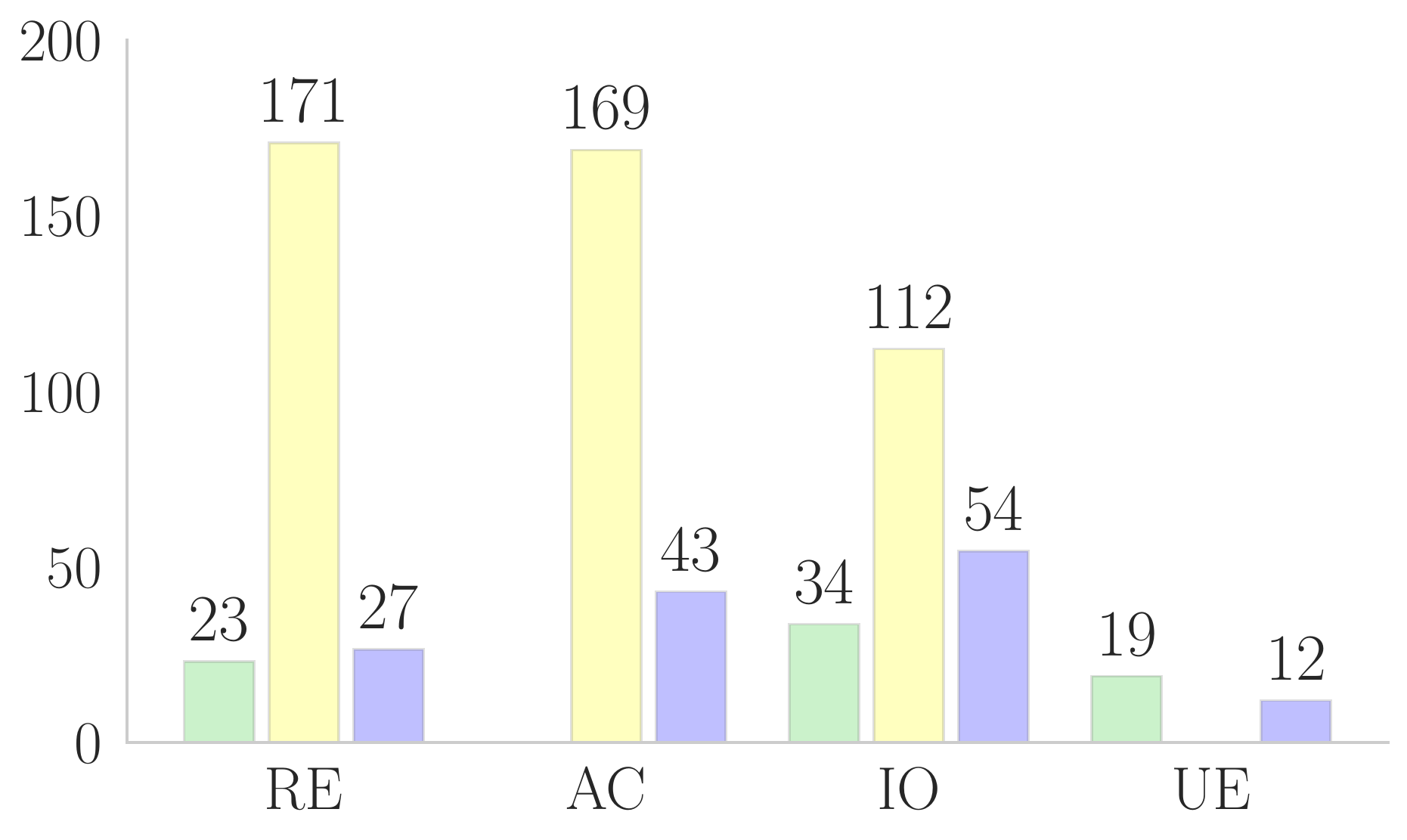}
         \caption{\textsc{SmartBugs}}
         \label{fig:smartbugsdeploymentcosts}
     \end{subfigure}
     \hfill
     \begin{subfigure}[b]{0.29\textwidth}
         \centering
         \includegraphics[width=1.0\columnwidth]{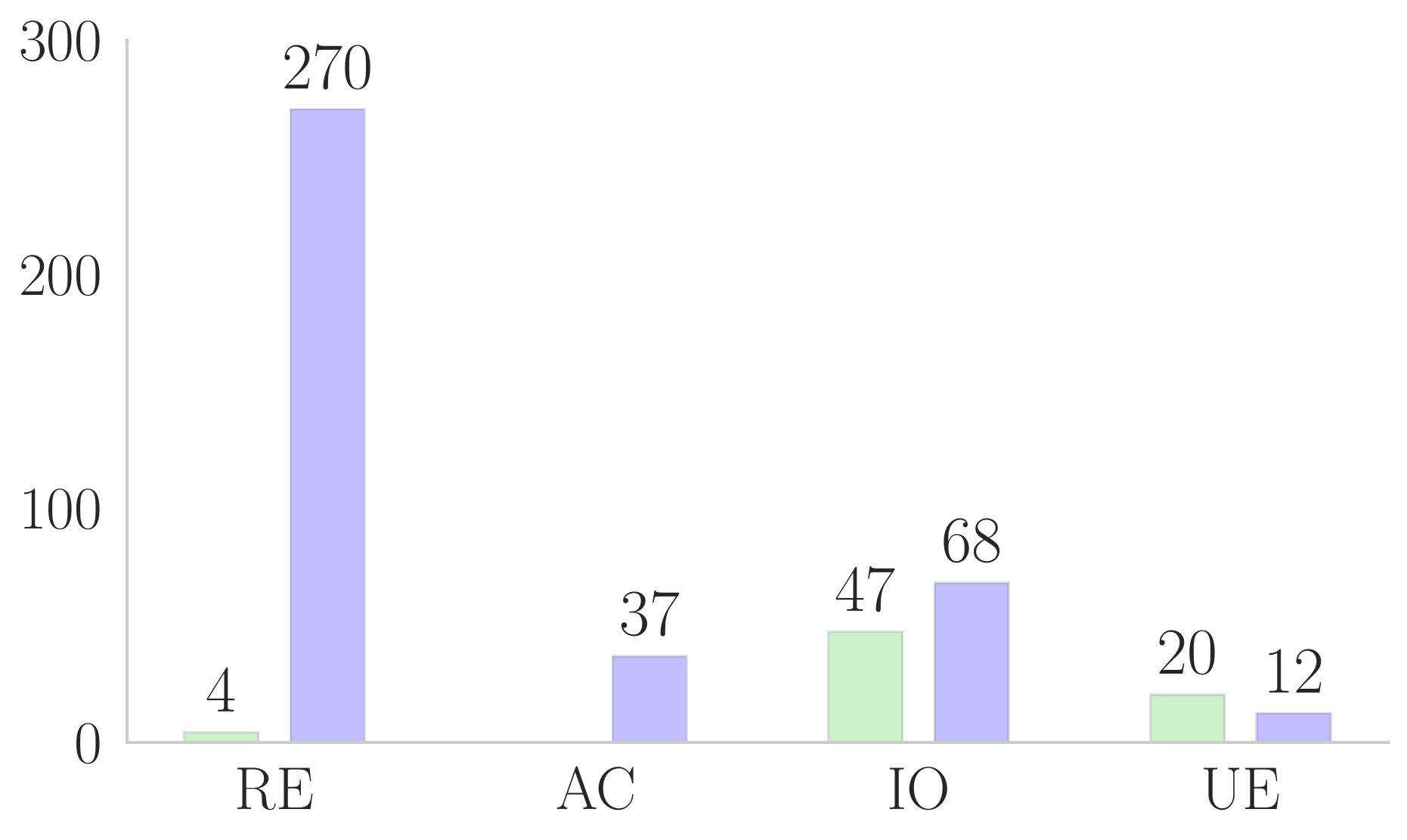}
         \caption{\textsc{Horus}}
         \label{fig:horusdeploymentcosts}
     \end{subfigure}
        \caption{Deployment cost increase in terms of bytes.}
        \label{fig:deploymentcosts}
\end{figure*}

\begin{figure*}
     \centering
     \begin{subfigure}[b]{0.35\textwidth}
         \centering
         \includegraphics[width=\textwidth]{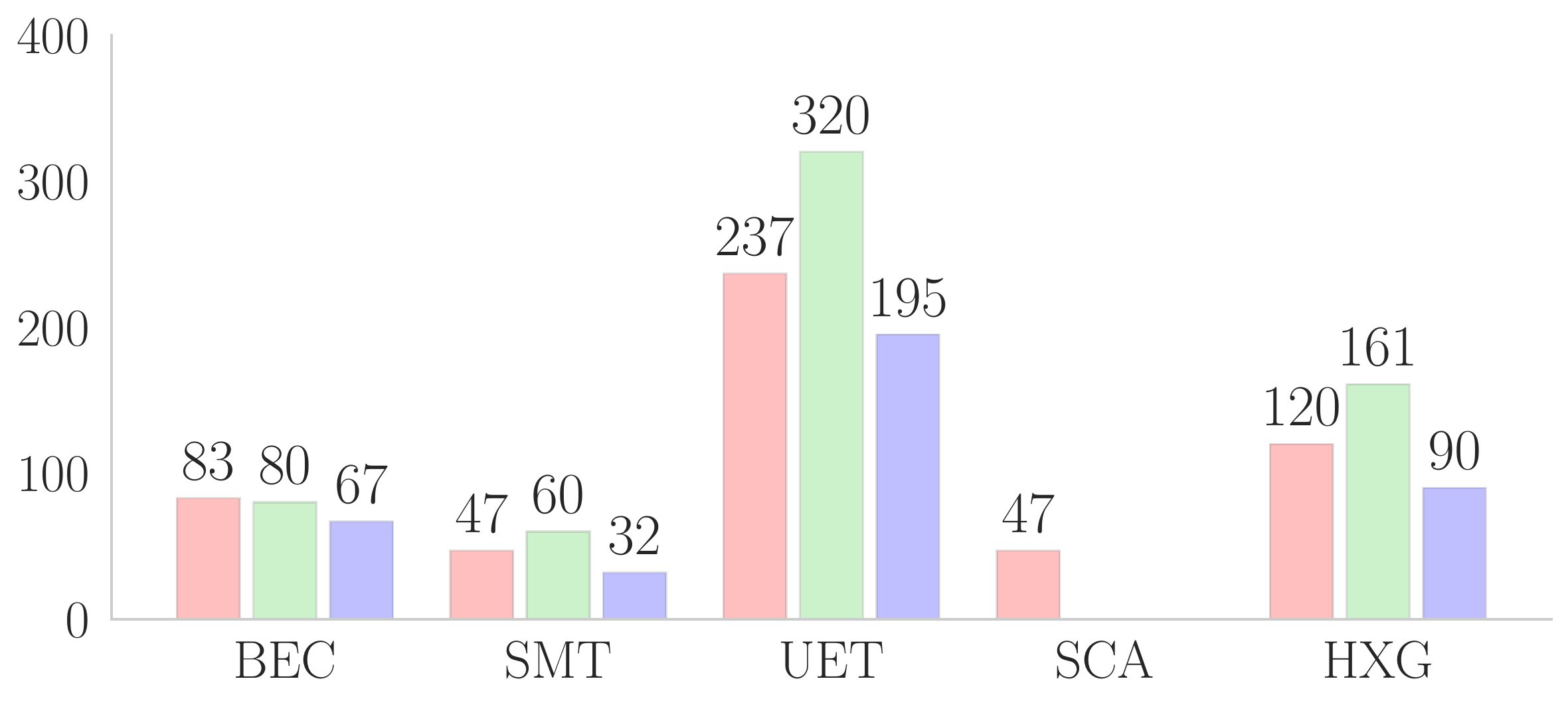}
         \caption{CVE}
         \label{fig:cvetransactioncosts}
     \end{subfigure}
     \begin{subfigure}[b]{0.35\textwidth}
         \centering
         \includegraphics[width=1.0\columnwidth]{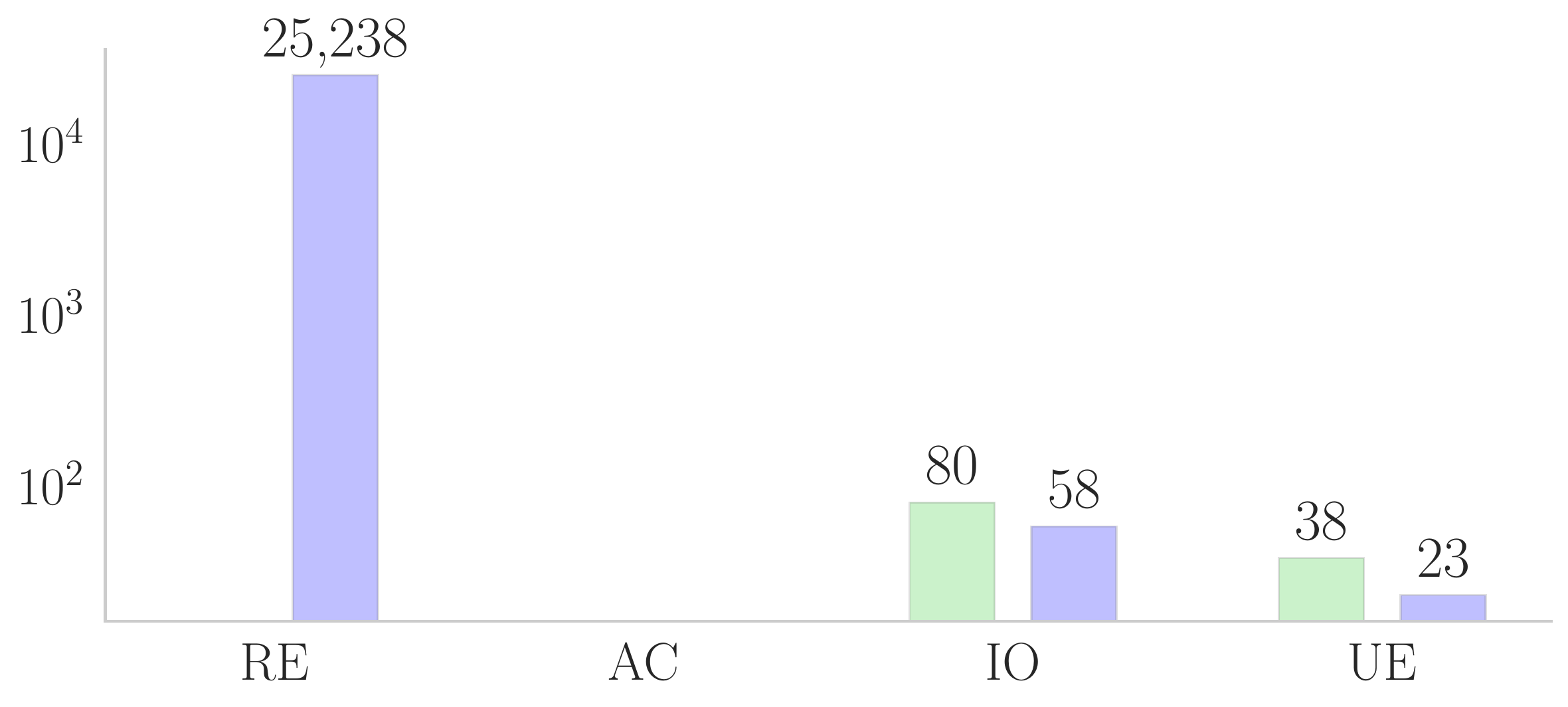}
         \caption{\textsc{Horus}}
         \label{fig:horustransactioncosts}
     \end{subfigure}
        \caption{Transaction cost increase in terms of gas.}
        \label{fig:transactioncosts}
        \vspace{0.2cm}
\end{figure*}

\vspace{0.2cm}
\noindent
\textbf{\textit{RQ3: Costs.}} We differentiate between \emph{deployment cost} and \emph{transaction cost}.
Deployment cost is associated to the cost when deploying a contract on the blockchain. It is computed based on the size of the bytecode. The larger the bytecode, the higher the cost.
Transaction cost (also known as runtime cost) is associated to the cost when executing a function of a smart contract. It is computed based on the gas consumed by the executed instructions. The more expensive instructions are executed, the higher the cost.
While deployment cost is a one-time cost, transaction cost is a repeating cost.
Our goal is therefore to primarily minimize transaction cost when introducing patches. 
\figurename~\ref{fig:deploymentcosts} highlights the deployment cost increase for all datasets. 
The deployment cost is measured by computing the difference in terms of size between the patched and the original bytecode.
We state that the patches introduced by \textsc{EVMPatch} and \textsc{sGuard} add the largest deployment cost. This is because those tools use templates that have been generated from source code.
In contrast, \toolname{} and \textsc{SmartShield}, use manually crafted and optimized bytecode level templates that use less instructions. 
\textsc{SmartShield} is in most cases the cheapest in terms of deployment cost.
This is because \textsc{SmartShield} injects its patching template only once into the code and then simply jumps to it. In addition, for reentrancy it does not introduce new code but rather tries to move the existing code around (\eg move writes to storage before a call). For example, \textsc{SmartShield} only adds on average 4 bytes of overhead for reentrancy on the \textsc{Horus} dataset, whereas \toolname{} adds 270 bytes. However, we have shown that \textsc{SmartShield}'s patching technique for reentrancy is often inaccurate. Moreover, for unhandled exceptions, \toolname{} is more efficient than \textsc{SmartShield}, because of its optimized patch template.
\figurename~\ref{fig:transactioncosts}, highlights the transaction cost increase measured for each dataset.
The transaction cost is measured by computing the gas usage difference between the patched and original contract for all successfully executed benign transactions.
We state that \toolname{} adds in almost all the cases the smallest overhead in terms of transaction cost. 
For instance, in the \textsc{Horus} dataset, \toolname{} only adds on average 58 gas units of overhead when patching integer overflows, whereas \textsc{SmartShield} adds 80. However, we also see that \toolname{} adds 25,238 gas units on average when patching reentrancy, whereas \textsc{SmartShield} adds none. This is because \textsc{SmartShield} does not add new code, while \toolname{} adds two storage instructions which consume together already 25,000 gas units. 
\figurename~\ref{fig:executiontime}
shows the execution time of \toolname{} in proportion to the contract size for the \textsc{Horus} dataset. 
The contracts are sorted according to their execution time/contract size. We observe that 90\% of the contracts are patched in less than 9 seconds. The median is 0.07 seconds and the maximum is around 24 minutes. 
\toolname{} spends roughly 70\% of the execution time on the recovery of the CFG. We also state that the execution time grows linear to the size of a smart contract. 

\vspace{-0.1cm}
\subsection{Limitations}

\toolname{} highly depends on the ability to fully recover the CFG and correctly infer the context regarding integer bounds and free storage locations to be able to correctly patch the bytecode of a smart contract. Our preliminary experiments on CFG recovery show that we are able to recover 100\% of the CFG for 88\% of the cases and that we are able to correctly infer storage locations for all tested contracts. Moreover, during our evaluation we were able to block all attacks related to integer overflows, which means that we are able to correctly infer the integer bounds. 
However, to prevent breaking semantics or introducing new bugs when patching, \toolname{} first checks if the CFG has been fully recovered by checking if the CFG contains any basic blocks that are unreachable (\ie basic blocks with no jump instructions pointing to them as well as no jump instructions pointing from them to existing basic blocks). If there are any unreachable basic blocks, then \toolname{} outputs a warning regarding the possibility of breaking the semantics of the smart contract and requests for the user's consent to continue.

Another limitation is \toolname{}'s evaluation.
There exist multiple techniques to validate the correctness of patches such that they do not only fix the bug but also do not introduce new bugs. Our evaluation follows the same strategy as previous works (\eg \cite{rodler2020evmpatch,zhang2020smartshield}) by re-executing previous transactions of real-world contracts on the patched bytecode. We split previous transactions in benign and attacking transactions. If an attacking transaction is blocked, then we assume that the patch is working correctly. If the result of a benign transaction is the same for the original bytecode and the patched bytecode, then we assume that the semantics have been preserved. However, this does guarantee soundness since our evaluation depends on previous inputs generated by users where it can still be the case that new bugs have been introduced or that the actual semantics have been changed while those previous inputs simply do not cover those new cases. An alternative could be to apply differential fuzzing on the original and patched version of the bytecode to detect discrepancies. The input generation could be even driven by symbolic execution that leverages a constraint solver to synthesize inputs for the fuzzer. However, this approach would also heavily depend on correctly inferring the CFG.

\section{Related Work}

Framing code patching as a search and optimization problem has led several
authors~\cite{weimer2009automatically,qi2014strength} to leverage well-established
heuristics and search algorithms to patch smart contacts.
\textsc{SCRepair}~\cite{yu2020smart} uses a genetic algorithm to find a patch.
There are  inherent limits in terms of quality and depth of the results. For instance, complex
reentrancy patterns, such as cross-function reentrancy or faulty access control, cannot
be trivially patched and contrary to claims made by \textsc{SCRepair}, patches linked
to transaction order dependency are not addressed. Moreover, genetic algorithms are
notoriously slow since a population of solutions needs to be evolved and this process
is entirely random.
Several techniques from automated program repair research have been applied to smart
contracts. Nguyen et al.~\cite{nguyen2021sguard} present a tool called \textsc{sGuard},
that patches smart contract vulnerabilities at the source code level.
The disadvantage of this approach is that
the compiler often adds unnecessary/unoptimized code, increasing bytecode size
and thus causing increased deployment and transaction costs. The main difference
with our work is that we patch directly at the bytecode level and can highly optimize our patches. Moreover, our tool is language independent, while
\textsc{sGuard} only works for Solidity. 
Recently,
the academic community has shifted its interest to automated patching of EVM bytecode. 
For instance, Ayoade et al.~\cite{ayoade2019defense} patch integer overflows via bytecode rewriting and verify the equivalence of orginal and patched contract via the Coq theorem prover. However, their verification does not scale and their approach is not context sensitive and therefore can not be used to patch reentrancy or access control.
Rodler et al. propose \textsc{EVMPatch}~\cite{rodler2020evmpatch}, a methodology that can patch integer overflow
and access control bugs at bytecode level.
Integer overflows are patched through hard-coded patches restricted to type
{\small \texttt{uint256}} overflows and underflows. In order to patch access control
patterns, the developer is required to use a custom domain-specific language for
specifying a contract specific patch. Thus, patching is not fully automated and the
developer is required to understand and fix the bug manually. Claims that unhandled
exceptions can be patched are not backed by experiments and patching access control
bugs (such as suicidal contracts and leaking contracts), requires manual effort for every contract. Our approach is fully automated, covers more classes of
bugs, and does not require the kind of manual preparation reported
in~\cite{rodler2020evmpatch}.  
Targeting more complex bugs, Zhang et al.~\cite{zhang2020smartshield} presented
\textsc{SmartShield}, which automatically patches integer overflows, reentrancy
bugs, and unhandled exceptions at the bytecode level. The tool is limited
to only use hard-coded patches for integer overflows of type {\small \texttt{uint256}}.
We observed in our experiments that \textsc{SmartShield} has issues in patching
reentrancy bugs due the complexity of identifying data and control dependencies
across bytecode. Our approach addresses these challenges by leveraging taint
analysis at the bytecode level to infer contract related information (\eg integer
bounds and free storage space) and uses it to generate automatically contract specific patches.
\section{Conclusion}

In this work, we presented \toolname{}, a tool to automatically patch smart contracts using context-related information that is inferred at the bytecode level.  
\toolname{} is currently able to patch 7 types of vulnerabilities. 
It can easily be extended by adding further vulnerability detectors and by writing new patch templates using our custom DSL.
We compared \toolname{} to existing tools by patching almost 2,000 smart contracts and replaying more than 500K transactions.
Our results show that \toolname{} is able to effectively and correctly patch at least 30\% more contracts than existing tools. 
Moreover, when compared to existing tools, the resulting transaction overhead is reduced by up to a factor of 1.7. 
We leave it to future work, to further optimize the overhead in terms of deployment costs.

\begin{acks}
We would like to thank our anonymous reviewers and our shepherd Kevin Roundy for their valuable comments and feedback. We also
gratefully acknowledge the support from the RIPPLE University Blockchain Research Initiative (UBRI) and the Luxembourg National Research Fund (FNR) under the grant 13192291.
\end{acks}

\bibliographystyle{ACM-Reference-Format}
\balance
\bibliography{references}

\end{document}